\documentstyle[12pt,aaspp,flushrt]{article}
\def\abs#1{\vert #1\vert}
\def\tends{\rightarrow}
\def\cf{{\it cf.\ }}
\def\be{\begin{equation}}
\def\ee{\end{equation}}

\def\bfm{{\bf m}}
\def\bfs{{\bf s}}
\def\bfr{{\bf r}}
\def\bfR{{\bf R}}
\def\bfk{{\bf k}}

\def\bfD{{\bf D}}
\def\bfP{{\bf P}}
\def\bfS{{\bf S}}
\def\half{{\textstyle{1\over2}}}
\def\bfI{{\bf I}}
\def\bfl{{\bf l}}
\def\bfr{{\bf r}}

\def\bfR{{\bf R}}

\def\bfv{{\bf v}}

\def\bfz{{\bf z}}
\def\bfW{{\bf W}}
\def\bfZ{{\bf Z}}

\def\bfw {{\bf w}}
\def\bfOmega {{\bf \Omega}}

\def\bnabla{\mbox{\boldmath $\nabla$}}
\def\df{{\caps df}}
\def\dfs{{\caps df}s}

\def\bfPsi{\mbox{\boldmath $\Psi$}}
\def\bfLambda{\mbox{\boldmath $\Lambda$}}
\def\half{{\textstyle{1\over2}}}

\def\erf{{\rm erf}}

\def\bfr{{\bf r}}
\def\bfx{{\bf x}}

\def\bfnabla{{\mbox{\boldmath $\nabla$}}}
\def\bfr{{\bf r}}

\def\bfF{{\bf F}}

\def\bfZ{{\bf Z}}

\def\bfOmega{{\bf \Omega}}
\def\bfk{{\bf k}}

\def\bfl{{\bf l}}
\def\bfv{{\bf v}}

\def\bfw{{\bf w}}
\def\bfI{{\bf I}}
\def\bfF{{\bf F}}

\def\Respw{{R(\bfr,\bfr^\prime,\omega)}}
\def\Respwp{{R(\bfr,\bfr^\prime,\omega^\prime)}}

\newfont{\caps}{cmcsc10}

\tighten

\begin{document}
\title{Linear response, dynamical friction and the
fluctuation-dissipation theorem in stellar dynamics}

\author{Robert W. Nelson$^{1,3}$ and Scott Tremaine$^{2,3}$}

\affil{$^1$Theoretical Astrophysics, California Institute of Technology 
130-33, Pasadena, California 91125, USA}

\medskip

\affil{$^2$Canadian Institute for Advanced Research,\\
Program in Cosmology and Gravity}

\medskip

\affil{$^3$Canadian Institute for Theoretical Astrophysics, McLennan Labs, 
University of Toronto, 60~St.\ George St.,
Toronto M5S 3H8, Canada}

\medskip
\begin{abstract}

\noindent
We apply linear response theory to a general, inhomogeneous,
stationary stellar system, with particular emphasis on dissipative
processes analogous to Landau damping.  Assuming only that the
response is causal, we show that the irreversible work done by an
external perturber is described by the anti-Hermitian part of a linear
response operator, and damping of collective modes is described by the
anti-Hermitian part of a related polarization operator. We derive an
exact formal expression for the response operator, which is the
classical analog of a well-known result in quantum statistical
physics. When the self-gravity of the response can be ignored, and the
ensemble-averaged gravitational potential is integrable, 
the expressions for the mode energy, damping rate, and polarization operator 
reduce to well-known formulae derived from
perturbation theory in action-angle variables. In this
approximation, dissipation occurs only via resonant interaction with
stellar orbits or collective modes.  For stellar systems in thermal
equilibrium, the anti-Hermitian part of the response operator is
directly related to the correlation function of the fluctuations. Thus
dissipative properties of the system are completely determined by the
spectrum of density fluctuations---the fluctuation-dissipation
theorem.  
In particular, we express the coefficient of dynamical
friction for an orbiting test particle in terms of the
fluctuation spectrum; this reduces to the known Chandrasekhar formula 
in the restrictive case of an infinite homogeneous system
with a Maxwellian velocity distribution.

\end{abstract}


\section{Introduction}

\label{sec:intro}

\noindent
The aim of this paper is to apply to stellar dynamics two of the powerful
tools that have been developed in other branches of statistical physics:
linear response theory and the fluctuation-dissipation theorem. 

The application of the methods of statistical physics to self-gravitating
stellar systems is challenging for several reasons: the interparticle forces
are long-range, the systems are intrinsically inhomogeneous, and true
thermodynamic equilibrium does not exist (e.g. Lynden-Bell \& Wood
1968). Thus, in practice, most of our understanding of relaxation in stellar
systems comes from analyses based on the following related approximations
(Jeans 1913, 1916; Chandrasekhar 1942; Binney \& Tremaine 1987; Spitzer 1987):

\begin{enumerate}

\item{Local approximation:} The stellar system is assumed to be infinite and
homogeneous, and the force field from the equilibrium system is
neglected; thus the unperturbed stellar orbits are straight lines at constant
velocity. The validity of this approximation requires that most of the
relaxation arises from close encounters, which have impact parameter $b\ll R$
where $R$ is the size of the system.  The failure of the local approximation
at large impact parameter is why the Coulomb logarithm is poorly determined in
stellar systems. 

\item{Markov approximation:} The interactions between stars are treated as
sequential binary encounters of negligible duration, so that the evolution of
the one-particle distribution function (hereafter \df) can be treated as a
Markov process. This is a reasonable approximation for short-range forces such
as those in neutral gases, but more questionable in a stellar system because
of the long range of the gravitational force. The Markov approximation is
closely related to the local approximation because only local encounters can
plausibly be considered to have negligible duration.

\item{Diffusion approximation:} Most of the relaxation is assumed to arise
from weak encounters, in which the orbital deflection is small; in practice
this requires $b\gg b_0\equiv Gm/\sigma^2$ where $m$ and $\sigma$ 
are a typical stellar
mass and velocity. In this approximation, the evolution of the stellar orbits
is a diffusion process that can be described by a Fokker-Planck equation. The
diffusion approximation is consistent with the local approximation if $b_0\ll
R$, so that most encounters are weak ($b\gg b_0$) while still local ($b\ll R$);
in this case the contribution to the relaxation from encounters with $b<b_0$
is smaller than the contribution from $b>b_0$ by of order the Coulomb
logarithm $\ln\Lambda=\ln R/b_0$. With the approximations listed so far, the
effects of relaxation on a stellar orbit are completely described by a
dynamical friction force, which gives the mean change in velocity per unit
time $\langle\Delta v_i\rangle$, and a diffusion tensor, which gives the
mean-square change in velocity per unit time $\langle\Delta v_i\Delta
v_j\rangle$.

\item{Neglect of self-gravity:} The self-gravity of the response of the
stellar system is neglected. Thus, for example, dynamical friction arises from
the gravitational drag exerted on a body by the wake it creates, but most
calculations of dynamical friction neglect the effects of the self-gravity of
the wake in determining its amplitude and shape. The error introduced by this
approximation can be significant: Weinberg (1989) and Hernquist \& Weinberg
(1989) find that including the self-gravity of the wake can suppress dynamical
friction on an orbiting satellite by a factor of 2--3. Errors of this
magnitude are often excused because the precise value of the Coulomb logarithm
is not known in any case; however, if the orbital frequency of the perturbing
body is close to the frequency of a collective mode of the system, then
the effects of self-gravity are increased dramatically (e.g. Weinberg
1993).

\end{enumerate}

The goal of this paper is to summarize some of the insights into fluctuations
and dissipation in stellar systems that can be derived {\it without} making
any of these approximations. Our tools will be linear response theory and the
fluctuation-dissipation theorem 
(e.g. Martin 1968, Forster 1975, Landau \& Lifshitz 1980, Reichl 1980,
Sitenko 1982, Klimontovich 1986, Kubo et al. 1991).

The intimate relation between fluctuation and dissipation in stellar systems
was first recognized by Chandrasekhar (1943), who argued that a star's random
walk in velocity due to stochastic gravitational forces must be balanced by a
drag force (dynamical friction) if the stochastic process is to leave a
Maxwellian distribution invariant. The most general form of this relation is
described by the fluctuation-dissipation theorem. This theorem relates the
fluctuations in a dynamical system in thermal equilibrium, described by the
correlation function, to the rate at which the system absorbs energy from a
weak external field, described by the response operator.  In the context of
stellar systems, the theorem relates the dynamical friction force to the
diffusion tensor (\cf eq. \ref{eq:dragc} below); however, we shall see that
this relation is far more general than the usual derivation of either the
dynamical friction force or the diffusion tensor individually.

The conceptual importance of the relation between dynamical friction and
stochastic forces was recognized by Bekenstein \& Maoz (1992) and Maoz (1993),
who concentrated on providing a unified derivation of both effects in a
homogeneous system; in
contrast, we shall focus on the general properties of drag and fluctuations
that can be derived without reference to specific systems.

We begin \S \ref{sec:linear} with a review of linear response theory
applied to a stationary inhomogeneous system, stressing the
constraints that causality places on the analytic properties of the
response operator; for example, we derive the Kramers-Kronig relations
between the Hermitian and anti-Hermitian parts of this operator and
show that dissipation in the system is determined by the
anti-Hermitian part of the response operator. We also define the
polarization and dielectric operators, and derive their relation to
the response operator. We employ these operators to examine collective
modes of the system in \S \ref{sec:coll}, deriving an expression for
the energy of a mode in terms of the Hermitian part of the
polarization operator, and show that its (Landau) damping rate is
determined by the anti-Hermitian part of the polarization operator.
In \S \ref{sec:hamiltonian} we derive an exact formal expression for
the response operator in a general Hamiltonian system, which has a
direct analog in quantum systems.  For systems described by integrable
potentials, we give an explicit form for the polarization operator in
terms of the one-particle \df.  We describe fluctuations in the system
using the correlation function in \S \ref{sec:fluct}, deriving the
symmetries that are imposed on the correlation function by the
principle of microscopic reversibility.  
In \S \ref{sec:thermal} we prove
the fluctuation-dissipation theorem, which relates the response
operator and the correlation function for an isothermal
system. We describe fluctuations in non-isothermal systems and the 
dressed-particle approximation in \S \ref{sec:dressed}. 
In \S \ref{sec:applications} we apply these results
to derive general relations between dynamical friction and relaxation
in isothermal stellar systems. 
We end with  a discussion in \S \ref{sec:discussion}, and a summary of the
main formulae and results in \S \ref{sec:summary}.

Throughout this paper, we shall apply the term ``equilibrium'' to a stellar
system when the one-particle \df\ is a solution of the time-independent
collisionless Boltzmann equation; such a system is stationary except for slow
evolution due to relaxation. We reserve the term ``thermal equilibrium'' for a
system whose $N$-particle \df\ is an exponential function of the $N$-particle
Hamiltonian. Most of our results apply to any equilibrium stellar system,
although some apply only to time-reversible systems (in which the Hamiltonian
and the equilibrium one-particle \df\ are invariant under time-reversal),
to systems in which the Hamiltonian is integrable,  or to
systems in thermal equilibrium.

\section{Linear response theory}
\label{sec:linear}

\noindent
In this section we examine the response of an equilibrium stellar system to
small potential perturbations, assuming that the induced density depends
linearly on the strength of the perturbation through an---as yet
unspecified---response operator. It is often difficult or impossible to find
an analytic expression for the response operator. Nevertheless, many of its
properties follow directly from causality and linearity.

\subsection{The response operator}

\label{sec:respop}

\noindent
We consider an equilibrium stellar system that is subjected to a small
perturbation $\Phi_e(\bfr,t)$ from an external potential.  The density
perturbation induced in the system, $\rho_s(\bfr,t)$, can be expressed in
terms of a linear response operator $R(\bfr,\bfr^\prime,\tau)$, defined by 
\be
\rho_s(\bfr,t)=\int d\bfr^\prime dt^\prime
R(\bfr,\bfr^\prime,t-t^\prime)\Phi_e(\bfr^\prime,t^\prime);
\label{eq:rdef}
\ee
causality requires $R(\bfr,\bfr^\prime,\tau)=0$ for $\tau<0$. 
We stress that $\Phi_e$ does {\it not} include the gravitational potential
arising from the response density $\rho_s$; in this respect the response
operator can be contrasted with the polarization operator defined in
\S\ref{sec:polar} below.

In this section we consider the general properties of the response operator,
deferring the derivation of explicit forms for $R(\bfr,\bfr^\prime,\tau)$
until \S \ref{sec:hamiltonian}.  We start by taking Fourier transforms, which
we denote by replacing the variable $\tau$ by $\omega$,
e.g. $\Phi_e(\bfr,\tau)=\int_{-\infty}^\infty
d\omega\Phi_e(\bfr,\omega)\exp(-i\omega t)$.  Equation (\ref{eq:rdef})
simplifies to 
\be \rho_s(\bfr,\omega)=2\pi \int d\bfr^\prime
R(\bfr,\bfr^\prime,\omega)\Phi_e(\bfr^\prime,\omega).
\label{eq:rrddff}  
\ee 
Since $R(\bfr,\bfr^\prime,\tau)$ is real, we must have \be
R^\ast(\bfr,\bfr^\prime,\omega)=R(\bfr,\bfr^\prime,-\omega).
\label{eq:symme}
\ee 
We can analytically continue the response operator so that it is defined for
complex frequencies $z$ (\cf eq. \ref{eq:respw} with $\omega \to z$).
Causality requires that $R(\bfr,\bfr^\prime,z)$ has
no poles in the upper half plane. 

Equation (\ref{eq:rrddff}) can be written as an operator equation,
\be
\rho_s=2\pi\bfR(\omega)\Phi_e.
\label{eq:rdefop}
\ee
If we define the inner product as
\be
(\psi,\phi)\equiv \int d\bfr \psi^\ast(\bfr)\phi(\bfr)d\bfr,
\label{eq:innerp}
\ee
then the adjoint operator $\bfR^\dagger(\omega)$ satisfies
\be
(\bfR^\dagger \psi,\phi)=(\psi,\bfR\phi)\quad\hbox{for all }
\psi(\bfr),\phi(\bfr),
\ee
and is given by
\be
R^\dagger(\bfr,\bfr',\omega)=R^\ast(\bfr',\bfr,\omega).
\ee 

We can always write the response operator in the form
\be
R(\bfr,\bfr^\prime,\omega)\equiv R_H(\bfr,\bfr^\prime,\omega) + 
R_A(\bfr,\bfr^\prime,\omega),
\ee
defined by 
\begin{eqnarray}
R_H(\bfr,\bfr^\prime,\omega) & \equiv & \half R(\bfr,\bfr^\prime,\omega) + 
\half R^\ast(\bfr^\prime,\bfr,\omega), \nonumber \\
R_A(\bfr,\bfr^\prime,\omega) & \equiv & \half R(\bfr,\bfr^\prime,\omega) - 
\half R^\ast(\bfr^\prime,\bfr,\omega).
\label{eq:hermdef}
\end{eqnarray} 
The operators $R_H$ and $R_A$ satisfy the relations
\be
R_H^\dagger(\bfr,\bfr^\prime,\omega)= 
R_H^\ast(\bfr^\prime,\bfr,\omega)  =  R_H(\bfr,\bfr',\omega), \quad
R_A^\dagger(\bfr,\bfr^\prime,\omega) =R_A^\ast(\bfr',\bfr,\omega) 
=  -R_A(\bfr,\bfr',\omega).
\label{eq:realexp}
\ee
Thus $R_H$ is Hermitian and $R_A$ is anti-Hermitian. Using equations
(\ref{eq:symme}) and (\ref{eq:hermdef}) it is easy to show that
\be
R_H^\ast(\bfr,\bfr^\prime,\omega)=R_H(\bfr,\bfr^\prime,-\omega),\qquad
R_A^\ast(\bfr,\bfr^\prime,\omega)=R_A(\bfr,\bfr^\prime,-\omega).
\label{eq:symmet}
\ee
 
In terms of time rather than frequency, we may write
\be
R(\bfr,\bfr^\prime,\tau)=R_e(\bfr,\bfr^\prime,\tau)+R_o(\bfr,\bfr^\prime,\tau),
\ee
where 
\be
R_e(\bfr,\bfr^\prime,\tau)=\int d\omega 
R_H(\bfr,\bfr^\prime,\omega)e^{-i\omega\tau},\quad 
R_o(\bfr,\bfr^\prime,\tau)=\int d\omega
R_A(\bfr,\bfr^\prime,\omega)e^{-i\omega\tau}.
\label{eq:rodef}
\ee
It is straightforward to show using (\ref{eq:symmet}) that $R_e$ and $R_o$ are
real. Using (\ref{eq:realexp}) we can show that 
\be
R_e(\bfr,\bfr^\prime,\tau)=R_e(\bfr^\prime,\bfr,-\tau),\qquad  
R_o(\bfr,\bfr^\prime,\tau)=-R_o(\bfr^\prime,\bfr,-\tau).
\label{eq:timesym}
\ee
Moreover, since $R(\bfr,\bfr^\prime,-\tau)=0$ for $\tau > 0$ we must have
\be
R_e(\bfr,\bfr^\prime,\tau)=R_o(\bfr,\bfr^\prime,\tau), \qquad 
R_e(\bfr,\bfr^\prime,-\tau)=-R_o(\bfr,\bfr^\prime,-\tau), \qquad \tau > 0.
\label{eq:kkt}
\ee

If $R(\bfr,\bfr^\prime,\tau)$ is well-behaved then $R(\bfr,\bfr^\prime,z)\to
0$ as Im$(z)\to\infty$, where $z$ is the complex frequency. Since 
$R(\bfr,\bfr^\prime,z)$ has no poles in the upper half plane, we can write
\be
\int_{-\infty}^{\infty} d\omega^\prime{\Respwp \over 
\omega^\prime-\omega+i\eta}=0,
\ee
where $\eta>0$, since we can close the contour at ${\rm Im}(z)=+\infty$.
We next make use of the identity
\be
\lim_{\eta\to0}{1\over x-y+i\eta}={\cal P}\left(1\over
x-y\right)-i\pi\,\hbox{sgn}(\eta)\delta(x-y),
\label{eq:lbk}
\ee
where $\cal P$ denotes the Cauchy principal value and $\delta$ denotes the
Dirac delta function. Then we have
\be
\Respw=
-{i\over \pi} \int_{-\infty}^\infty
 d\omega^\prime \Respwp {\cal P}\left(1\over \omega^\prime-\omega\right),
\label{eq:respw}
\ee
Equating the Hermitian and anti-Hermitian components,
\begin{eqnarray}
R_H(\bfr,\bfr^\prime,\omega) & = & 
-{i\over \pi} \int_{-\infty}^\infty 
d\omega^\prime R_A(\bfr,\bfr^\prime,\omega^\prime){\cal P}\left(1
\over \omega^\prime-\omega\right), \nonumber \\
R_A(\bfr,\bfr^\prime,\omega) 
& = & -{i\over \pi}
\int_{-\infty}^\infty d\omega^\prime R_H(\bfr,\bfr^\prime,\omega^\prime)
{\cal P}\left(1\over \omega^\prime-\omega\right).
\label{eq:kramers}
\end{eqnarray}
These are the Kramers-Kronig relations, which follow from causality and 
require no other assumptions about the dynamics of the stellar system.
They can be analytically continued to complex frequencies $z=\omega+i\eta$. 

Another relation is obtained by examining the integral
\be
\int{z\over z^2+s^2}R(\bfr,\bfr',z)dz,
\ee
where $s$ is real and positive, and the integral is taken along the real axis
and closed in the upper half-plane (e.g., Landau \& Lifshitz 1980). Since
$R(\bfr,\bfr',z)$ has no poles in the upper half-plane, the only contribution
to the integral comes from the pole at $z=is$; thus
\be
R(\bfr,\bfr',is)=-{i\over\pi}\int_{-\infty}^\infty {\omega
R(\bfr,\bfr',\omega) d\omega\over \omega^2+s^2}.
\ee
Equating the Hermitian and anti-Hermitian components,
\begin{eqnarray}
R_H(\bfr,\bfr^\prime,is) & = & 
-{i\over \pi} \int_{-\infty}^\infty 
d\omega {\omega R_A(\bfr,\bfr^\prime,\omega)\over \omega^2+s^2}, \nonumber \\
R_A(\bfr,\bfr^\prime,is) & = & -{i\over \pi}
\int_{-\infty}^\infty d\omega {\omega R_H(\bfr,\bfr^\prime,\omega)
\over \omega^2+s^2}. 
\end{eqnarray}

\subsection{The polarization and dielectric operators}

\label{sec:polar}

\noindent
It is sometimes useful to introduce a different measure of the linear 
response: the polarization operator relates the induced density to the 
{\it total} potential produced by an external perturbation,
$\Phi_{t}=\Phi_e+\Phi_s$,
\be
\rho_s(\bfr,t)=
\int d\bfr^\prime dt^\prime P(\bfr,\bfr^\prime,t-t^\prime)
\Phi_t(\bfr^\prime,t^\prime),
\label{eq:P}
\ee 
where $\Phi_s$ is associated with the induced density $\rho_s$ by
Poisson's equation, $\nabla^2\Phi_s=4\pi G\rho_s$.  In frequency space
$P(\bfr,\bfr^\prime,\omega)$ has the same analytic properties described in \S
\ref{sec:respop} for $R(\bfr,\bfr^\prime,\omega)$; for example,
$P(\bfr,\bfr^\prime,\omega)$ satisfies Kramers-Kronig relations. In general
$P$ is easier to compute explicitly than $R$ because it does not depend on the
self-gravity of the response, which is already included in $\Phi_t$.

It is straightforward to show that the two operators are related by the
following nonlinear integral equations, 
\begin{eqnarray}
R(\bfr,\bfr',\omega)& = & P(\bfr,\bfr',\omega)+2\pi\int d\bfx d\bfx'
P(\bfr,\bfx,\omega)\Psi(\bfx,\bfx')R(\bfx',\bfr',\omega),\nonumber \\
& = & P(\bfr,\bfr',\omega)+2\pi\int d\bfx d\bfx'
R(\bfr,\bfx,\omega)\Psi(\bfx,\bfx')P(\bfx',\bfr',\omega),
\label{eq:rp}
\end{eqnarray}
where
\be
\Psi(\bfx,\bfx')=-{G\over|\bfx-\bfx'|}
\label{eq:coul}
\ee
is the Coulomb interaction potential. These can be written in operator
notation as
\be
\bfR(z)=\bfP(z)+2\pi\bfP(z)\bfPsi\bfR(z)=
\bfP(z)+2\pi\bfR(z)\bfPsi\bfP(z).
\label{eq:rprp}
\ee
Note that $\bfPsi$ is self-adjoint, $\bfPsi^\dagger=\bfPsi$. 

Poisson's equation may be written
\be
\Phi_s(\bfr,\omega)=\int d\bfx\Psi(\bfr,\bfx)\rho_s(\bfx,\omega)= 2\pi\int
d\bfx d\bfr'\Psi(\bfr,\bfx)P(\bfx,\bfr',\omega)\Phi_t(\bfr',\omega).
\ee
Thus the external potential is related to the total potential by 
\be
\Phi_e(\bfr,\omega)=\int d\bfr'D(\bfr,\bfr',\omega)\Phi_t(\bfr',\omega),
\label{eq:didef}
\ee
where
\be
D(\bfr,\bfr^\prime,\omega)=\delta(\bfr-\bfr^\prime)-
2\pi \int d\bfx \Psi(\bfr,\bfx)P(\bfx,\bfr^\prime,\omega)
\label{eq:dielectric}
\ee
is the dielectric response operator. The name arises by analogy to the
dielectric constant, which similarly relates the total electric field inside a
dielectric to the field generated by external sources. In operator notation
\be
\Phi_e=\bfD(z)\Phi_t\quad\hbox{where}\quad 
\bfD(z)=\bfI-2\pi\bfPsi\bfP(z).
\label{eq:ddop}
\ee
The inverse operator satisfies
\be
\Phi_t=\bfD^{-1}(z)\Phi_e,\qquad
\bfD^{-1}(z)=\bfI+2\pi\bfPsi\bfR(z).
\ee
This is easily verified by computing $\bfD\bfD^{-1}$ and using equation
(\ref{eq:rprp}). 
The density that generates the external potential is related to the total
density by 
\be
\rho_t=\bfLambda(z)\rho_e\quad\hbox{where}\quad 
\bfLambda(z)=\bfI+2\pi\bfR(z)\bfPsi, 
\label{eq:ldop}
\ee
and $\Phi_e=\bfPsi\rho_e$.
Its inverse is 
\be
\rho_e=\bfLambda^{-1}(z)\rho_t\quad\hbox{where}\quad 
\bfLambda^{-1}(z)=\bfI-2\pi\bfP(z)\bfPsi.
\label{eq:ldopinv}
\ee
The operators $\bfD^{-1}(z)$ and $\bfLambda(z)$ are singular when the response
operator $\bfR(z)$ is singular, which occurs at the frequencies of the
collective modes of the stellar system (\S \ref{sec:coll}). 

Two further identities can be derived using (\ref{eq:rprp}),
\be
\bfP(z)=\bfR(z)\bfD(z),
\label{eq:qqid}
\ee 
\be
\bfR(z)=\bfLambda(z)\bfP(z).
\label{eq:qqida}
\ee 

The operators $\bfR$, $\bfP$, $\bfD$ and $\bfLambda$ 
generally do not commute, and are 
related through the nonlinear integral equations above.
These expressions simplify, however, for an infinite homogeneous
stellar system. In this case, the operators are translationally invariant: 
the spatial dependence of $D(\bfr,\bfr^\prime,\omega)$, for
example, enters only through 
$\bfr-\bfr^\prime$. Thus we can define a spatial Fourier transform, 
\be
D(\bfk,\omega)=
\int d(\bfr-\bfr^\prime) D(\bfr-\bfr^\prime,\omega) 
e^{-i\bfk\cdot(\bfr-\bfr^\prime)},
\ee
so that operating with $\bfR$, $\bfP$, $\bfD$ or $\bfLambda$ reduces to 
scalar multiplication by a function of the wavelength 
$\bfk$ and the wavenumber $\omega$.
In particular for the gravitational Coulomb potential given by 
(\ref{eq:coul}), 
\be
\Psi(\bfk)=-{4\pi G \over k^2}.
\ee
Then, 
\be
D(\bfk,\omega)=1+{8\pi^2 G\over k^2}P(\bfk,\omega), 
\ee
\be
R(\bfk,\omega)={P(\bfk,\omega)\over D(\bfk,\omega)}
\ee
\be
\Lambda(\bfk,\omega)={R(\bfk,\omega)\over P(\bfk,\omega)}
={1\over D(\bfk,\omega)}.
\ee
Similar expressions for an infinite homogeneous electron plasma follow with 
the substitution
$Gm^2 \tends -e^2$, where $-e$ is the charge of the electron.

\subsection{Work done by an external potential}
\label{sec:work}

\noindent
We gain insight into the response operator by considering the work
done on the stellar system by an external potential $\Phi_e(\bfr,t)$;
we assume for simplicity that $\Phi_e\to 0$ as $t\to\pm\infty$. 
The rate at which the external potential does work on a unit mass is
$-\bnabla\Phi_e\cdot\bfv$, where $\bfv$ is the velocity; thus the rate
of doing work on the stellar system is
\begin{eqnarray}
\dot E & = & -\int d\bfr \bnabla\Phi_e\cdot\rho\bfv,\nonumber \\
       & = &  \int d\bfr \Phi_e\bnabla\cdot(\rho\bfv),\nonumber \\
       & = & -\int d\bfr \Phi_e(\bfr,t){\partial\rho(\bfr,t)\over\partial t},
\label{eq:edot}
\end{eqnarray}
where the last line follows from the continuity equation. Thus the
total energy change is
\be
\Delta E_s =\int \dot E dt = -\int d\bfr\int dt
\Phi_e(\bfr,t){\partial\rho(\bfr,t)\over\partial t}  = \int d\bfr\int dt
\rho(\bfr,t){\partial\Phi_e(\bfr,t)\over\partial t}.
\ee
Using the relation
\be
\int_{-\infty}^\infty A^\ast(t)B(t)dt=2\pi\int_{-\infty}^\infty
A^\ast(\omega)B(\omega)d\omega
\label{eq:wiener}
\ee
and the definition of the response operator (\ref{eq:rrddff})  we have
\begin{eqnarray}
\Delta E_s & = & 2\pi i\int d\bfr \omega d\omega 
\rho(\bfr,\omega)\Phi_e^\ast(\bfr,\omega) \nonumber \\
 & = & (2\pi)^2i\int d\bfr d\bfr' \omega d\omega
R(\bfr,\bfr',\omega)\Phi_e(\bfr',\omega) \Phi_e^\ast(\bfr,\omega).
\end{eqnarray}
The contribution of the Hermitian component of the response operator
to this integral vanishes, so that
\begin{eqnarray}
\Delta E_s & = &(2\pi)^2i\int d\bfr d\bfr' \int_{-\infty}^\infty \omega d\omega
R_A(\bfr,\bfr',\omega)\Phi_e(\bfr',\omega)
\Phi_e^\ast(\bfr,\omega)\nonumber \\
 & = &8\pi^2i\int d\bfr d\bfr' \int_0^\infty \omega d\omega
R_A(\bfr,\bfr',\omega)\Phi_e(\bfr',\omega)
\Phi_e^\ast(\bfr,\omega).
\end{eqnarray}

An important special case of this formula occurs when the external
potential is nearly monochromatic, with frequency $\omega_0$. Then we
may write
\begin{eqnarray}
\Delta E_s & = &8\pi^2i\omega_0\int d\bfr d\bfr'R_A(\bfr,\bfr',\omega_0)
\int_0^\infty d\omega \Phi_e(\bfr',\omega)
\Phi_e^\ast(\bfr,\omega)\nonumber \\
 & = & 4\pi i\omega_0\int d\bfr d\bfr'R_A(\bfr,\bfr',\omega_0)
\int_{-\infty}^\infty dt \Phi_e(\bfr',t)\Phi_e^\ast(\bfr,t).
\end{eqnarray}
For example, suppose that the external potential has the form
\be 
\Phi_e(\bfr,t)=g(t){\rm Re}\left [{\phi_e(\bfr)}e^{-i\omega_0 t}\right ],
\label{eq:excite}
\ee where the amplitude $g(t)$ is assumed to be real, vanishingly
small in the distant past and future, and varies slowly in the sense
that $|\dot g/g|={\rm O}(\epsilon)\ll|\omega_0|$. Then the component of
$\Phi_e$ with frequency near $+\omega_0$ is $\half
g(t)\phi_e(\bfr)\exp(-i\omega_0t)$, and thus 
\be 
\Delta E_s = \pi
i\omega_0\int dt g^2(t)(\phi_e,\bfR_A\phi_e). 
\label{eq:deone} 
\ee

It is worthwhile to re-derive this result another way. We begin with
the last line of equation (\ref{eq:edot}). The induced density may be
written $\rho(\bfr,t)={\rm Re}[\rho_s(\bfr,t)e^{-i\omega_0 t}]$, where
\be 
\rho_s(\bfr,t)= \int d\bfr^\prime \int_0^\infty d\tau
R(\bfr,\bfr^\prime,\tau)\phi_e(\bfr^\prime) g(t-\tau)e^{i\omega_0\tau}.
\label{eq:rhos}
\ee
Since $g(t)$ changes slowly, we can expand
$g(t-\tau)=g(t)-\tau\dot g(t)+{\rm O}(\epsilon^2)$ 
in (\ref{eq:rhos}), so that 
\be
\rho_s(\bfr,t)=
2\pi g(t)\int d\bfr^\prime R(\bfr,\bfr^\prime,\omega_0)\phi_e(\bfr^\prime)
+ 2\pi i \dot g(t)\int d\bfr^\prime {\partial R\over \partial \omega}
(\bfr,\bfr^\prime,\omega_0) \phi_e(\bfr^\prime)+{\rm O}(\epsilon^2).
\ee
Thus
${\partial \rho/\partial t}={\rm Re}[(-i\omega_0 \rho_1+\dot\rho_2)
e^{-i\omega_0 t}]+{\rm O}(\epsilon^2)$, where 
\be
\rho_1(\bfr,t)=2\pi g(t) \int d\bfr^\prime R(\bfr,\bfr^\prime,\omega_0)
\phi_e(\bfr^\prime),\qquad \dot \rho_2(\bfr,t)=
2\pi\dot g(t) \int d\bfr^\prime {\partial \omega R \over \partial \omega}
(\bfr,\bfr^\prime,\omega_0)\phi_e(\bfr^\prime).
\label{eq:lasteq}
\ee
We average the rate of doing work over one cycle of the external 
potential, denoting this average by $\langle\cdot\rangle$. 
The time average of the integrand in (\ref{eq:edot}) can be written 
\be 
\left \langle \Phi_e{\partial\rho\over\partial t} \right \rangle
= -{i\omega_0\over 4}(\phi^*_e\rho_1- \phi_e\rho^*_1)+ {1\over 4}(\phi^*_e\dot
\rho_2 +\phi_e\dot \rho^*_2).
\label{eq:avg}
\ee
Finally, writing  $R$ in terms of its Hermitian and 
anti-Hermitian components, using equations (\ref{eq:realexp}) and 
(\ref{eq:lasteq}), we find
\be
\langle \dot E\rangle = W+\dot E_{\rm int},
\label{eq:absorb}
\ee
where
\be
W=\pi i\omega_0 \int d\bfr d\bfr^\prime
\phi_e^*(\bfr)R_A (\bfr,\bfr^\prime,\omega_0)
\phi_e(\bfr^\prime)=\pi i\omega_0(\phi_e,\bfR_A\phi_e),
\label{eq:damp}
\ee
\be
E_{\rm int} =  - {\pi\over 2}
\int d\bfr d\bfr^\prime
\phi_e^*(\bfr){\partial \omega R_H\over \partial \omega}
(\bfr,\bfr^\prime,\omega_0)\phi_e(\bfr^\prime)=-{\pi\over 2}
\left(\phi_e,{\partial\omega\bfR_H\over\partial \omega}\phi_e\right),
\label{eq:erg}
\ee
and $g(t)$ is taken to be unity. 

The quantity $W$ is related to the energy change derived earlier
through $\Delta E_s=\int W dt$; $W$ involves the anti-Hermitian part
of the response operator and is present even if the external potential is
maintained at constant amplitude ($g(t)=$constant). It
represents the rate of doing work on the stellar system that is required to
maintain the periodic potential; thus the anti-Hermitian response $R_A$ is 
associated with energy absorption (dissipation) by the stellar 
system\footnote{We refer to this process as energy absorption or
dissipation; however, in some collisionless systems energy can be emitted
rather than absorbed by this process.}. In collisionless stellar systems with
integrable potentials, the work done by the external potential is absorbed by
particle resonances or collective modes with frequencies near $\omega_0$ (\cf 
\S \ref{sec:action}).

The term $E_{\rm int}$ in (\ref{eq:absorb}) represents the work
required to build up the response to the time-varying field.  This
work must be done even when the dissipation associated with $R_A$ is
small or zero.  We can thus regard $E_{\rm int}$ as the total energy
associated with the interaction of the external field and the stellar
system. If the response operator $R_H$ is non-singular at the
perturbing frequency, this energy will be recovered by the external
perturber if the potential is turned off adiabatically.

\section{Collective modes}
\label{sec:coll}

\noindent
In the previous section we have considered the induced density response of a
stellar system that is subjected to an external potential. However, a stellar
system may also support a self-induced response even when no external
perturbation is present ($\rho_s\not=0$ even when
$\Phi_e=0$). Such a response is called a collective mode;
equation (\ref{eq:didef}) implies that the potential $\Phi_s$ associated with a
collective mode satisfies the dispersion relation
\be
\int d\bfr'D(\bfr,\bfr',z)\Phi_s(\bfr',z)=0\quad \hbox{or} \quad
\bfD(z)\Phi_s=0,
\label{eq:disp}
\ee
where $z$ is the complex frequency. The response operator $\bfR$ is singular
at the eigenfrequency of a collective mode (\cf eq. \ref{eq:rdefop} and
Kalnajs 1971).

For an infinite, homogeneous medium 
the collective modes are plane waves, $\Phi_s\propto\exp(i\bfk\cdot\bfr)$,
and equation (\ref{eq:disp}) reduces to an algebraic dispersion relation,
$D(\bfk,z)=0$.

\subsection{The mode energy}

\noindent
We can use the results of \S \ref{sec:work} to determine the energy of a
collective mode in the case where the imaginary component of the
eigenfrequency is small. We imagine perturbing the
stellar system with an external potential of the form (\ref{eq:excite}), where
$\omega_0$ is close to the real part of the eigenfrequency of the mode. 
Averaging equation
(\ref{eq:edot}) over one period $2\pi/\omega_0$ and writing the external
potential in terms of the total and response potentials, 
$\Phi_e=\Phi_t-\Phi_s$, we find
\be
\langle \dot E\rangle=-\int d\bfr
\left\langle\Phi_e{\partial\rho\over\partial t}\right\rangle
=\int d\bfr\left\langle\Phi_s{\partial\rho\over\partial t}\right\rangle
-\int d\bfr\left\langle\Phi_t{\partial\rho\over\partial
t}\right\rangle.
\ee
The first term can be written $dU_m/dt$ where 
\be
U_m={1\over 2}\int d\bfr\left \langle \rho\Phi_s \right \rangle =
{1\over 8}\left[(\rho_s,\phi_s)+(\phi_s,\rho_s)\right]
={\pi\over 4}\left[(\bfP\phi_t,\phi_s)+(\phi_s,\bfP\phi_t)\right]
\ee
is the potential energy of the induced disturbance.
The second term can be evaluated along the lines of 
equations (\ref{eq:rhos})--(\ref{eq:absorb}):
\be
-\int d\bfr\left\langle\Phi_t{\partial\rho\over\partial
t}\right\rangle = W_m+\dot K_m,
\ee
where
\be
W_m=\pi i\omega_0 \int d\bfr d\bfr^\prime
\phi_t^*(\bfr)P_A (\bfr,\bfr^\prime,\omega_0)
\phi_t(\bfr^\prime)=\pi i\omega_0(\phi_t,\bfP_A\phi_t),
\label{eq:dampb}
\ee
\be
K_m=-{\pi\over 2}\int d\bfr
d\bfr'\phi_t^\ast(\bfr){\partial\omega 
P_H\over\partial \omega}(\bfr,\bfr',\omega_0)\phi_t(\bfr')=-{\pi\over 2}\left(
\phi_t,{\partial\omega\bfP_H\over\partial\omega}\phi_t\right).
\ee
Since $\omega_0$ is close to the eigenfrequency of the mode, we expect that the
response of the system is strong, $|\Phi_s|\gg|\Phi_e|$. Thus to a good
approximation we can replace $\phi_t$ by $\phi_s=\phi_t-\phi_e$, 
and we find that the total 
energy associated with a collective mode is 
\be
E_m = K_m + U_m =-{\pi\over 2}\left(
\phi_s,{\partial\omega\bfP_H\over\partial\omega}\phi_s\right)+
{\pi\over 2}(\phi_s,\bfP_H\phi_s)=-{\pi\over 2}\omega_0\left(
\phi_s,{\partial\bfP_H\over\partial\omega}\phi_s\right),
\ee
and the energy is dissipated at a rate
\be
W_m=\pi i\omega_0(\phi_s,\bfP_A\phi_s).
\label{eq:dampc}
\ee
Since $U_m$ is the potential energy and $E_m$ is the total energy, we may
identify $K_m$ as the kinetic energy of the mode. 

The quantity $W_m$ represents the rate of absorption of energy from the mode
by the stellar system. The amplitude of the mode varies as $\exp(\eta t)$,
where $\eta$ is the imaginary part of the eigenfrequency; thus  
\be
\eta={1\over 2E_m}{dE_m\over dt}=-{W_m\over 2E_m}=
{ i(\phi_s,\bfP_A\phi_s)\over(\phi_s,\bfP_{H,\omega}\phi_s)},
\label{eq:landa}
\ee
where $\bfP_{H,\omega}=\partial\bfP_H/\partial\omega$. 
These results are only valid in the limit of weak damping,
$|\eta|\ll|\omega_0|$, since we have assumed that the eigenfrequency of the
mode is close to the real frequency $\omega_0$.

Equation (\ref{eq:landa}) can also be derived directly from the
dispersion relation. Taking the inner product of (\ref{eq:disp}) with the
density of the mode gives
\be
(\rho_s,\bfD(z_0)\phi_s)=0,
\ee
where $z_0=\omega_0+i\eta$ is the complex eigenfrequency of the mode. 
We write $D$ in terms of its Hermitian and anti-Hermitian parts, and assume
that $\eta$ is small so that we can expand to first order in $\eta$,
\be
(\rho_s,\bfD_H(\omega_0)\phi_s)+(\rho_s,\bfD_A(\omega_0)\phi_s)
+i\eta(\rho_s,\bfD_{H,\omega}(\omega_0)\phi_s)=0.
\ee
It is straightforward to show that the first term is real and 
the remaining terms are purely imaginary.
Equating the imaginary parts to zero and solving for $\eta$, we recover
(\ref{eq:landa}).

\section{The response operator for a Hamiltonian system}
\label{sec:hamiltonian}

\noindent
We now examine the dynamics of the stellar system in more detail.
We consider a system composed of $N$ stars, with phase-space
coordinates $\bfz_i=(\bfr_i,\bfv_i)$, $i=1,\ldots,N$ (note that we define
phase space using velocity, not momentum). For simplicity we assume
that all of the stars are identical, with mass $m$, although the results we
derive can be generalized to a range of stellar masses. We
denote the coordinates and Hamiltonian of the system in $6N$-dimensional phase
space by $\bfZ\equiv(\bfz_1,\ldots,\bfz_N)$ and $mH_0(\bfZ)$, where
\be
H_0(\bfZ)=\half\sum_{i=1}^N\bfv_i^2+\half m\sum_{i\neq j}
\Psi(\bfr_i-\bfr_j);
\label{eq:hamgen}
\ee 
here $\Psi$ is given by equation (\ref{eq:coul}).

We consider an ensemble of stellar systems, described by a $N$-particle
distribution function (\df) $f(\bfZ,t)$, where $f$ is a symmetric
function of the $N$ variables $\bfz_1,\ldots,\bfz_N$, normalized so that $\int
f(\bfZ,t)d\bfZ=1$. Thus the ensemble average of any phase function $u(\bfZ)$
may be written $\langle u \rangle=\int u(\bfZ)f(\bfZ,t)d\bfZ$. The evolution
of the \df\ is described by Liouville's equation 
\be 
{\partial f\over\partial t}+[f,H_0]=0, 
\label{eq:liouville}
\ee 
where $[\ ,\ ]$ is a Poisson bracket.

The trajectory of the system 
$\tilde\bfZ(t)$ satisfies Hamilton's equations
\be
{d\tilde\bfZ\over dt}=[\bfZ,H_0]_{\tilde\bfZ}.
\label{eq:hamdef}
\ee 
We shall use $\bfZ_\tau$ as shorthand for the image of $\bfZ$ after time
$\tau$ under the Hamiltonian flow (\ref{eq:hamdef}); thus, if $\tilde\bfZ(t)$
is a trajectory and $\bfZ=\tilde\bfZ(t_0)$, then
$\bfZ_\tau=\tilde\bfZ(t_0+\tau)$.

We may derive a generalized response operator for two arbitrary 
phase functions $u_\bfs(\bfZ)$, $v_\bfs(\bfZ)$, where 
$\bfs$ is a parameter that labels these functions. 
Imagine that we apply a small external perturbation $mH_1(\bfZ,t)$ to the
stellar system. We suppose that $mH_1$ can be written in the form
\be
mH_1(\bfZ,t)=\int d\bfs^\prime v_{\bfs^\prime}(\bfZ)X(\bfs^\prime,t).
\label{eq:hextdef}
\ee 
The induced
perturbation in the expectation of the phase function $u_\bfs$ may then 
be written
\be 
u_{1\bfs}(t)=\int d\bfs^\prime dt^\prime
R_{uv}(\bfs,\bfs^\prime,t-t^\prime)X(\bfs^\prime,t^\prime),
\label{eq:ressdef}
\ee
where $R_{uv}(\bfs,\bfs^\prime,t-t^\prime)$ is a generalized 
response operator (\cf eq. \ref{eq:rdef}). 

The perturbation to the \df\ induced by $H_1$ is written
$f_1(\bfZ,t)$ and satisfies the linearized Liouville equation,
\be
{df_1\over dt}={\partial f_1\over\partial t}+[f_1,H_0]=-[f_0,H_1],
\ee
where $d/dt$ denotes the Lagrangian derivative along the unperturbed
trajectory. The formal solution to this equation is
\be
f_1(\bfZ,t)=-\int_0^\infty d\tau[f_0,H_1(t-\tau)]_{\bfZ_{-\tau}}.
\ee 
Thus we may write
\begin{eqnarray}
u_{1\bfs}(t) & = & \int d\bfZ f_1(\bfZ,t)u_\bfs(\bfZ) \nonumber \\
   & = & -\int d\bfZ u_\bfs(\bfZ)
          \int_0^\infty d\tau[f_0,H_1(t-\tau)]_{\bfZ_{-\tau}} \nonumber \\
   & = & -{1\over m}\int d\bfZ u_\bfs(\bfZ)\int d\bfs^\prime 
 \int_0^\infty d\tau X(\bfs^\prime,t-\tau)[f_0,v_{\bfs^\prime}]_{\bfZ_{-\tau}}.
\end{eqnarray}
Thus the response operator is
\be
R_{uv}(\bfs,\bfs^\prime,\tau)=-{\Theta(\tau)\over m}\int d\bfZ
u_\bfs(\bfZ)[f_0,v_{\bfs^\prime}]_{\bfZ_{-\tau}},
\label{eq:uuiioo}
\ee
where $\Theta$ is the step function. 
Using the identity $A[B,C] = [AB,C] - B[A,C]$ this can be simplified to
\begin{eqnarray}
R_{uv}(\bfs,\bfs^\prime,\tau) & = & {\Theta(\tau)\over m}\int d\bfZ
f_0(\bfZ_{-\tau})[u_\bfs(\bfZ),v_{\bfs^\prime}(\bfZ_{-\tau})]\nonumber \\
& = &  {\Theta(\tau)\over m}\int d\bfZ
f_0(\bfZ)[u_\bfs(\bfZ_\tau),v_{\bfs^\prime}(\bfZ)]\nonumber \\
& = & {\Theta(\tau)\over m}\langle [u_\bfs(t+\tau),v_{\bfs^\prime}(t)] \rangle.
\label{eq:ruv}
\end{eqnarray}
That is, the response operator is just the ensemble-averaged Poisson bracket
of $u_\bfs(t+\tau)$ and $v_{\bfs^\prime}(t)$. 

The density-density response operator is defined by setting
$X(\bfr,t)=\Phi_e(\bfr,t)$, where $\Phi_e$ is the external perturbing
potential, and $u_\bfr(t) = v_{\bfr}(t)=\rho(\bfr,t)$ where
\be
\rho(\bfr,t)=m\sum_{i=1}^N\delta[\bfr_i(t)-\bfr],
\label{eq:lllkkjj}
\ee
is the exact density distribution for the $N$-body 
system with coordinates $\bfz_i(t)=(\bfr_i(t),\dot\bfr_i(t))$. Thus
\be
R(\bfr,\bfr^\prime,\tau)
= {\Theta(\tau)\over m}\langle [\rho(\bfr,t+\tau),\rho(\bfr^\prime,t)] \rangle.
\label{eq:respdens}
\ee 
The Poisson bracket is taken with respect to the phase-space coordinates
$\bfZ=(\bfz_1,\ldots,\bfz_N)$ at time $t$, which determine $\bfr_i(t)$ and
$\bfr_i(t+\tau)$ and hence implicitly determine $\rho(\bfr,t+\tau)$ and
$\rho(\bfr^\prime,t)$ through (\ref{eq:lllkkjj}).

A nearly identical result to (\ref{eq:ruv}) was first derived by 
Kubo (1957) in the context of quantum statistical mechanics.
In this case the $N$-body \df\ is replaced by the 
quantum mechanical density operator, which satisfies an equation of
motion very similar to Liouville's equation (\ref{eq:liouville});
the primary difference being that the Poisson bracket is replaced by a 
quantum mechanical commutator.

The expression (\ref{eq:respdens}) for the response operator is formally
exact, but difficult or impossible to evaluate in practice for realistic 
stellar
systems.  However, the analogous expression for the polarization operator is
simpler. In this case we examine the response of the system to a total
potential, and do not have to consider the additional gravitational
forces arising from perturbations to the stellar orbits. Thus the 
Hamiltonian (\ref{eq:hamgen}) can be written as
\be
H_0(\bfZ)=\half\sum_{i=1}^N\bfv_i^2+ \sum_{i}
\Phi_i(\bfr_i,t);
\label{eq:hamgenp}
\ee 
here $\Phi_i(\bfr,t)=m\sum_{j\not=i}\Psi[\bfr-\bfr_{0j}(t)]$ 
is the potential from all the stars other than $i$, moving along their 
unperturbed orbits. This Hamiltonian is separable,
$H_0(\bfZ)=\sum_{i=1}^NH_i(\bfz_i)$; the responses of different stars are
independent and hence we can work in 6-dimensional phase space instead of
$6N$-dimensional phase space. There is a further simplification in 
the limit of large $N$, where the potentials
$\Phi_i$ can be replaced by the ensemble-averaged potential
$\Phi_0$ (the ``mean-field'' approximation, \cf \S \ref{sec:thermal}), 
and the analog to equation (\ref{eq:respdens}) is
\be
P(\bfr,\bfr^\prime,\tau)
= \Theta(\tau)\int d\bfx d\bfv F_0(\bfx,\bfv)
[\delta(\bfx_\tau(\bfx,\bfv)-\bfr),\delta(\bfx-\bfr')].
\label{eq:pespdens}
\ee
Here $F_0(\bfx,\bfv)$ is the one-particle \df, defined so that 
$F_0(\bfx,\bfv)d\bfx d\bfv$ is the
(ensemble-averaged) mass in the phase-space volume $d\bfx d\bfv$ (in contrast
to the $N$-particle \df, the integral of $F_0$ over phase space is normalized
to the total mass rather than to unity). The Poisson bracket is taken with
respect to $\bfx,\bfv$; and $\bfx_\tau(\bfx,\bfv)$ is the position at time 
$\tau$ of the particle that was at $(\bfx,\bfv)$ at time 0.

\subsection{Symmetries of the response operator}
\label{sec:symmetries}

\noindent
For particular systems, the response and polarization operators
may have symmetries in addition to those discussed in \S \ref{sec:respop}. 

A common situation is that the equilibrium stellar system is invariant under
time reversal (e.g. non-rotating galaxies). Under time reversal stellar
positions remain the same, while velocities are reversed, $(\bfr_i,\bfv_i)
\tends (\bfr_i,-\bfv_i)$. The density transforms as
$\rho(\bfr,t) \tends m\sum_i\delta(\bfr-\bfr_i(-t)) = \rho(\bfr,-t)$, while
the Poisson bracket reverses sign.  Consequently, under time reversal
\begin{eqnarray}
\langle [\rho(\bfr,t+\tau),\rho(\bfr^\prime,t)] \rangle & \tends &
-  \langle [\rho(\bfr,-t-\tau),\rho(\bfr^\prime,-t)] \rangle
\nonumber \\
& = &  - \langle [\rho(\bfr,t),\rho(\bfr^\prime,t+\tau)] \rangle
\nonumber \\
& = & \langle [\rho(\bfr^\prime,t+\tau),\rho(\bfr,t)] \rangle.
\end{eqnarray}
Comparing to equation (\ref{eq:respdens}), for systems invariant under time
reversal we must have 
\be R(\bfr,\bfr^\prime,\tau)=R(\bfr^\prime,\bfr,\tau)
\ee 
This result implies in turn that the Hermitian component of the response
operator $R_H(\bfr,\bfr',\omega)$ is real 
and even in $\omega$, while the anti-Hermitian component
$R_A(\bfr,\bfr',\omega)$ is imaginary and odd in $\omega$.
Similar relations can be proved
for the polarization operator $P(\bfr,\bfr',\tau)$, starting from equation
(\ref{eq:pespdens}).

Phase functions generally transform under time reversal
as $u[\tilde\bfZ(t)] \tends \epsilon^T_u u[\tilde\bfZ(-t)]$, 
where $\epsilon^T_u=+1$ for mass or energy density, while
$\epsilon^T_u=-1$ for momentum or angular momentum density
(e.g. Martin 1968). For systems invariant under time reversal, 
the general response function satisfies, 
\be
R_{uv}(\bfr,\bfr^\prime,\tau)= \epsilon^T_u \epsilon^T_v R_{vu}(\bfr^\prime,\bfr,\tau).
\ee 
Likewise for stellar systems invarient under parity transformations
($\bfr_i \tends -\bfr_i$), 
\be
R_{uv}(\bfr,\bfr^\prime,\tau)= \epsilon^P_u \epsilon^P_v 
R_{uv}(-\bfr,-\bfr^\prime,\tau).
\ee 
where $\epsilon_u^P$ is the signature of $u(\bfZ)$ under parity
transformation.

\subsection{Goodman's stability criterion}
\label{sec:reverse}

\noindent
The results of the previous section can be used to re-derive an
elegant instability test for a time-reversible stellar system
(Goodman 1988). Consider the operator
\be
\bfW(z)=\bfP(z)-{1\over 2\pi}\bfPsi^{-1};
\ee
where $\bfPsi^{-1}=\nabla^2/(4\pi G)$ (\cf eq. \ref{eq:coul}). A collective
mode $\phi_s$ satisfies $\bfW(z)\phi_s=0$. Assume that the frequency $z=is$
where $s$ is real and positive. Then $\bfP(is)$ is real, and hence Hermitian
if the stellar system is time-reversible. Similarly $\bfW$ is real and
Hermitian. Therefore the eigenvalues $\lambda(s)$ of $\bfW(is)$ are
real. Moreover as $s\to \infty$, $\bfP(is)\to 0$ so $\bfW\to
-(2\pi)^{-1}\bfPsi^{-1}$, which is positive-definite for the inner product
(\ref{eq:innerp}). Now let $s$ decrease from infinity. If for any $s_0>0$
there is a function $\phi$ such that 
\be 
(\phi,\bfW(is_0)\phi)<0, 
\ee
then $\bfW(is_0)$ is no longer positive-definite. Thus some of its eigenvalues
$\lambda(s_0)$ are negative. Therefore there is some $s_1>s_0$ for which one
of the eigenvalues is zero. Hence there is an unstable mode with growth rate
$s_1$. In other words a necessary condition for stability in time-reversible
stellar systems is that
\be
2\pi(\phi,\bfP(is_0)\phi)<(\phi,\bfPsi^{-1}\phi)
\ee
for all $\phi$ and for all $s_0>0$. This is a slight generalization of
Goodman's result, which was derived only for systems with integrable
potentials. 

\subsection{The polarization operator in action-angle variables}
\label{sec:action}

\noindent
The polarization operator (\ref{eq:pespdens}) can be evaluated explicitly if
the potential $\Phi_0(\bfr)$ is regular, so that phase space can be described
by action-angle coordinates $(\bfI,\bfw)$. The orbits are given by
$\bfI=$constant, $\bfw=\bfOmega t+\bfw_0$ where $\bfOmega=\partial
H_0/\partial \bfI$, and $mH_0$ is the Hamiltonian corresponding to
$\Phi_0$. Jeans's theorem states that the one-particle
\df\ depends only on the actions,
$F_0=F_0(\bfI)$. The canonical transformation from $(\bfx,\bfv)$ to
$(\bfI,\bfw)$ conserves the volume element in phase space: $d\bfx d\bfv=d\bfw
d\bfI$.\footnote{We define the canonical momentum and the actions without the
usual factor $m$, so these variables have dimensions (velocity) and
(velocity)$\times$(length) respectively.}

Consider a single star of unit mass with action-angle
coordinates ($\bfI,\bfw$); the corresponding spatial coordinate is
$\bfx(\bfI,\bfw)$. Formally, the spatial density of the star can be written
\be 
p(\bfr\vert \bfI,\bfw)=\delta[\bfr-\bfx(\bfI,\bfw)].
\label{eq:projdef}  
\ee 
Since $\bfw$ is cyclic, we can expand this in a Fourier series, 
\be 
p(\bfr\vert \bfI,\bfw)=\sum_\bfl p_\bfl(\bfr\vert\bfI)e^{i\bfl\cdot\bfw}
\qquad \hbox{where} \qquad
p_\bfl(\bfr\vert\bfI)={1\over (2\pi)^3} \int d\bfw e^{-i\bfl\cdot\bfw}
p(\bfr\vert \bfI,\bfw).
\label{eq:projop}
\ee
The $p_\bfl$'s are projection operators which give the 
Fourier components of any function of position $g(\bfr)$ that is expanded in a
Fourier series of the form $\sum_\bfl g_\bfl(\bfI)\exp(i\bfl\cdot\bfw)$: 
\begin{eqnarray}
g_\bfl(\bfI) & = & {1\over (2\pi)^3}\int d\bfw
g[\bfx(\bfI,\bfw)]e^{-i\bfl\cdot\bfw} \nonumber \\
             & = & {1\over (2\pi)^3}\int d\bfw d\bfr g(\bfr)\delta[\bfr-
\bfx(\bfI,\bfw)]e^{-i\bfl\cdot\bfw} \nonumber \\
             & = & 
{1\over (2\pi)^3}\int d\bfw d\bfr g(\bfr)p(\bfr\vert\bfI,\bfw)
e^{-i\bfl\cdot\bfw} \nonumber \\
             & = & \int d\bfr p_\bfl(\bfr| \bfI)g(\bfr).
\label{eq:wweezz}
\end{eqnarray}
It is straightforward to prove the following useful identity: if $h(\bfI)$ is
any function of the actions,
\be
(2\pi)^3\sum_{\bfl}\int d\bfI h(\bfI)p_\bfl^\ast(\bfr|\bfI)p_\bfl(\bfr'|\bfI)=
\delta(\bfr-\bfr')\int d\bfv h[\bfI(\bfr,\bfv)].
\ee
In particular, if $h(\bfI)$ is the one-particle \df\ $F_0(\bfI)$, then
\be
(2\pi)^3\sum_{\bfl}\int d\bfI F_0(\bfI)
p_\bfl^\ast(\bfr|\bfI)p_\bfl(\bfr'|\bfI)=
\delta(\bfr-\bfr')\rho_0(\bfr).
\label{eq:orthog}
\ee

The polarization operator (\ref{eq:pespdens}) can now be written as
\begin{eqnarray}
P(\bfr,\bfr^\prime,\tau)
& = & \Theta(\tau)\int d\bfw d\bfI F_0(\bfI)\sum_{\bfl,\bfm}
\left[p_\bfl^\ast(\bfr\vert\bfI)e^{-i\bfl\cdot(\bfw+\bfOmega\tau)},
p_\bfm(\bfr'\vert\bfI)e^{i\bfm\cdot\bfw}\right]\nonumber \\
&= & -(2\pi)^3\Theta(\tau)i\int d\bfI F_0(\bfI)\sum_{\bfl}
\bfl\cdot{\partial\over\partial\bfI}\left[p_\bfl^\ast(\bfr\vert\bfI)
p_\bfl(\bfr'\vert\bfI)e^{-i\bfl\cdot\bfOmega\tau}\right]\nonumber \\
&= & (2\pi)^3\Theta(\tau)i\int d\bfI \sum_{\bfl}
\bfl\cdot{\partial F_0(\bfI)\over\partial \bfI} p_\bfl^\ast(\bfr\vert\bfI)
p_\bfl(\bfr'\vert\bfI)e^{-i\bfl\cdot\bfOmega\tau},
\label{eq:dddsss}
\end{eqnarray}
where the last line follows through integration by parts.

The Fourier transform of the polarization operator is
\be 
P(\bfr,\bfr^\prime,\omega)= (2\pi)^2 \sum_\bfl \int d\bfI {p_\bfl^*(\bfr\vert
\bfI)p_\bfl(\bfr^\prime \vert \bfI) \over
\bfl\cdot\bfOmega-i\epsilon-\omega}
\,\bfl\cdot {\partial F_0\over\partial \bfI},
\label{eq:resp}
\ee
where $\epsilon$ is a small positive number. This can be split into Hermitian
and anti-Hermitian components using the identity (\ref{eq:lbk}):
\be
P(\bfr,\bfr^\prime,\omega)\equiv P_H(\bfr,\bfr^\prime,\omega) + 
P_A(\bfr,\bfr^\prime,\omega),
\ee
where (\cf eq. \ref{eq:lbk})
\begin{eqnarray}
P_H(\bfr,\bfr^\prime,\omega) & \equiv &
(2\pi)^2\sum_\bfl \int d\bfI\, 
p_\bfl^*(\bfr\vert \bfI)p_\bfl(\bfr^\prime \vert \bfI) 
\bfl\cdot{\partial F_0\over\partial \bfI}\,
{\cal P}\left({1\over \bfl\cdot \bfOmega-\omega}\right),\nonumber \\ 
P_A(\bfr,\bfr^\prime,\omega) & \equiv &
4\pi^3i\sum_\bfl \int d\bfI\, 
p_\bfl^*(\bfr\vert \bfI)p_\bfl(\bfr^\prime \vert \bfI) 
\bfl\cdot{\partial F_0\over\partial \bfI}\,
\delta(\bfl\cdot \bfOmega-\omega).
\label{eq:resptwo}
\end{eqnarray}
Thus $P_A$ is determined entirely by resonant stars satisfying
$\bfl\cdot\bfOmega=\omega$.

If the one-particle \df\ depends only on the energy per unit mass 
$E$, $F_0=F_0(E)$, then these expressions are simplified; for example,
\be
P_A(\bfr,\bfr^\prime,\omega) = 4\pi^3i\omega \sum_\bfl \int d\bfI 
p_\bfl^*(\bfr\vert \bfI)p_\bfl(\bfr^\prime \vert \bfI) 
{dF_0\over dE}\delta(\bfl\cdot \bfOmega-\omega).
\label{eq:isoth}
\ee

Equation (\ref{eq:resptwo}) allows us to find an explicit expression 
for the rate of energy dissipation in a collective mode 
(eq. \ref{eq:dampc}),
\be
W_s = -4\pi^4 \omega \sum_\bfl \int d\bfI\, 
\abs{\phi_\bfl(\bfI)}^2 \bfl\cdot{\partial F_0\over\partial \bfI}\,
\delta(\bfl\cdot \bfOmega-\omega),
\ee
where the potential of the collective mode has been expanded as
$\phi_s(\bfr)=\sum_\bfl \phi_\bfl(\bfI)\exp(i\bfl\cdot\bfw)$, and we have used
(\ref{eq:wweezz}).  This result is closely related to formulae originally
given by Lynden-Bell \& Kalnajs (1972) and can also be derived using
second-order Lagrangian perturbation theory (Nelson \& Tremaine 1995).
Similar expressions to those in this subsection are also given by
Tremaine \& Weinberg (1984), Goodman (1988), and Palmer (1994).

\section{Fluctuations}
\label{sec:fluct}

\noindent
The density and potential of any stellar system fluctuate about their mean
local values due to the finite number of stars. These fluctuations are
described by the density-density correlation function and its Fourier
transform.  For an isothermal stellar system, it turns out that the
correlation function is directly related to the response function. Thus,
dissipation processes described by $\bfR_A(\omega)$ are determined by the
fluctuation spectrum of the stellar system---the fluctuation-dissipation
theorem.

\subsection{The correlation function}

\noindent
We examine the density fluctuations $\delta\rho(\bfr,t)$ in the stellar
system, 
\be
\delta \rho(\bfr,t) = m \sum_i^N \delta[\bfr_i(t)-\bfr]-\rho_0(\bfr),
\ee
where $\rho_0(\bfr)= mN \langle \delta(\bfr_i-\bfr) \rangle$ is the 
mean density at $\bfr$
and as usual $\langle\cdot\rangle$ denotes an ensemble average.  
The density fluctuations are characterized by the correlation function
\be
C(\bfr,\bfr^\prime,\tau)=
\langle\delta\rho(\bfr,t+\tau) 
\delta\rho(\bfr^\prime,t)\rangle.
\label{eq:ccdef}
\ee
Since the equilibrium system is stationary, the correlation function is
independent of time $t$. If we replace $t$ by $t-\tau$ we derive the
symmetry relation
\be
C(\bfr,\bfr^\prime,\tau)=C(\bfr^\prime,\bfr,-\tau).
\label{eq:trev}
\ee

The Fourier transform of the correlation function is called the 
dynamic form factor: 
\be 
S(\bfr,\bfr^\prime,\omega)= {1\over 2\pi}\int d\tau e^{i\omega \tau} 
C(\bfr,\bfr^\prime,\tau) = {1\over 2\pi}\int d\tau e^{i\omega \tau} 
\langle\delta\rho(\bfr,t+\tau) \delta\rho(\bfr^\prime,t)\rangle, 
\label{eq:fform}
\ee 
Alternatively, we can define $S(\bfr,\bfr^\prime,\omega)$ 
by expressing the correlation function in terms
of the frequency transform of the density fluctuations, 
\be
C(\bfr,\bfr^\prime,\tau)=\int d\omega 
e^{-i\omega \tau} \int d\omega^\prime 
\langle\delta\rho(\bfr,\omega) 
\delta\rho(\bfr^\prime,\omega^\prime)\rangle
e^{-i(\omega+\omega^\prime)t}.
\ee
In order that the integral on the right side be independent of 
time $t$, we must have
\be
\langle\delta\rho(\bfr,\omega) \delta\rho(\bfr^\prime,-\omega^\prime)\rangle
=S(\bfr,\bfr^\prime,\omega)\delta(\omega-\omega^\prime).
\label{eq:rrrnnn}
\ee

The dynamic form factor satisfies the symmetry relations
\be
S^\ast(\bfr,\bfr^\prime,\omega)=S(\bfr,\bfr^\prime,-\omega),
\qquad S^\ast(\bfr,\bfr^\prime,\omega)=S(\bfr^\prime,\bfr,\omega);
\ee
the first of these follows because the correlation function is real, and the
second follows from (\ref{eq:trev}) and implies that the 
dynamic form factor is 
Hermitian. 
Finally, the correlation function at zero time difference is given 
by integrating $S(\bfr,\bfr^\prime,\omega)$ over all frequencies, 
\be
C(\bfr,\bfr^\prime,0)=\int d\omega S(\bfr,\bfr^\prime,\omega)
= \langle \delta \rho(\bfr,t)\delta \rho(\bfr^\prime,t) \rangle;
\label{eq:staform}
\ee
$C(\bfr,\bfr^\prime,0)$ is sometimes referred to as the static
form factor.

If the \df\ of the stellar system is invariant under time-reversal (\cf
\S\ref{sec:reverse}), and the Hamiltonian has the form (\ref{eq:hamgen}),
the correlation function satisfies the principle of microscopic reversibility,
\be
C(\bfr,\bfr^\prime,\tau)=C(\bfr^\prime,\bfr,\tau)\quad\hbox{or}\quad 
C(\bfr,\bfr^\prime,\tau)=C(\bfr,\bfr^\prime,-\tau).
\label{eq:micro}
\ee
This implies in turn that the dynamic form factor $S(\bfr,\bfr^\prime,\omega)$
is real, symmetric in $\bfr$ and $\bfr^\prime$, and an even function of
$\omega$.

A simple proof of (\ref{eq:micro}) relies on notation and results from
\S \ref{sec:hamiltonian}. We consider a generalized correlation function 
associated with any two phase functions $u_\bfs(\bfZ)$, $v_\bfs(\bfZ)$, 
labeled by the parameter $\bfs$. This may be written
\be
C_{uv}(\bfs,\bfs^\prime,\tau)=\langle\delta u_\bfs(t+\tau)
\delta v_{\bfs^\prime}(t)\rangle=
\int d\bfZ f_0(\bfZ)u_\bfs(\bfZ_\tau) v_{\bfs^\prime}(\bfZ)-\langle
u_\bfs\rangle \langle v_{\bfs^\prime}\rangle.
\label{eq:corrdef}
\ee

The density-density correlation function may be written in the form
\be
C(\bfr,\bfr^\prime,\tau)=
\int d\bfZ f_0(\bfZ)v_\bfr(\bfZ_\tau)v_{\bfr^\prime}(\bfZ)-
\rho_0(\bfr)\rho_0(\bfr^\prime),
\label{eq:corrdefa}
\ee where $v_\bfr(\bfZ)=\rho(\bfr,t)$ is defined by equation
(\ref{eq:lllkkjj}), and $\rho_0(\bfr)\equiv \langle v_\bfr(\bfZ)\rangle$ is
the ensemble-averaged density at $\bfr$. Now change the dummy variable from
$\bfZ$ to $\tilde\bfZ$, where $\tilde\bfZ$ is obtained from $\bfZ$ by
reversing all of the velocities. Since $f_0$ is time-reversible,
we can replace $f_0(\tilde\bfZ)$ by $f_0(\bfZ)$. We can replace
the volume element $d\tilde\bfZ$ by $d\bfZ$ because the Jacobian
$|\partial(\tilde\bfZ)/\partial(\bfZ)|=1$. Moreover,
$v_{\bfr^\prime}(\tilde\bfZ)=v_{\bfr^\prime}(\bfZ)$ because $v_\bfr$ depends
only on coordinates, not momenta. Finally,
$(\tilde\bfZ)_\tau=\widetilde{\bfZ_{-\tau}}$ because reversing the velocities
yields the time-reversed orbit in the Hamiltonian (\ref{eq:hamgen}); thus
$v_\bfr[(\tilde\bfZ)_\tau]=v_\bfr(\bfZ_{-\tau})$. Equation (\ref{eq:corrdefa})
becomes \be C(\bfr,\bfr^\prime,\tau)= \int d\bfZ
f_0(\bfZ)v_\bfr(\bfZ_{-\tau})v_{\bfr^\prime}(\bfZ)-
\rho_0(\bfr)\rho_0(\bfr^\prime)=C(\bfr,\bfr^\prime,-\tau),
\label{eq:corrdefb}
\ee
and equation (\ref{eq:micro}) then follows from (\ref{eq:trev}). 

For some purposes it is useful to define a modified correlation function that
removes long-term correlations:
\be
\widetilde C(\bfr,\bfr',\tau)=C(\bfr,\bfr',\tau)-C(\bfr,\bfr',\infty).
\label{eq:cmod}
\ee 
In contrast to most systems examined in statistical mechanics, the
correlation function does not vanish at large times in most stellar
systems. This is simple to understand in the context of spherical systems. The
time-averaged density of each star is an annulus with fixed orientation and
fixed inner and outer radii. The time-averaged total density is composed of
the sum of the densities from $N$ such annuli, and hence contains permanent
irregularities that are not present in the ensemble-averaged density. We
expect a non-zero correlation function at large times unless the system is
ergodic.

Correlations between particles are present for two conceptually distinct
reasons: random fluctuations in the number of particles in a given small
volume, which are present even if the particles move on their unperturbed
trajectories; and gravitational interactions between particles, which deflect
their mutual orbits.  It is sometimes useful to isolate the correlation
function arising only from random fluctuations in particle number, which we
denote 
\be 
C^{(0)}(\bfr,\bfr^\prime,\tau)=\langle\delta\rho(\bfr,t+\tau)
\delta\rho(\bfr^\prime,t)\rangle^{(0)}.
\label{eq:c0}
\ee
As in equation (\ref{eq:pespdens}), 
in the limit of large $N$ we can replace the exact potential
$\Phi_i$ by the ensemble-averaged potential $\Phi_0$, and write
the correlation function as
\be
C^{(0)}(\bfr,\bfr^\prime,\tau)
= m\int d\bfx d\bfv F_0(\bfx,\bfv)
\delta(\bfx_\tau(\bfx,\bfv)-\bfr)\delta(\bfx-\bfr')-{1\over
N}\rho_0(\bfr)\rho_0(\bfr'). 
\label{eq:cesdens}
\ee 
The factor proportional to 1/N derives from a near-cancellation of terms
analogous to the derivation of equation (\ref{eq:dampp}) below.

Because the particle orbits are independent in this approximation,
$\delta\rho(\bfr,t)$ is a Gaussian random field, and hence its properties are
completely described by the correlation function $C^{(0)}(\bfr,\bfr',\tau)$.

\subsection{The correlation function in action-angle variables}

\noindent
We can find an explicit expression for the density-density correlation
function $C^{(0)}(\bfr,\bfr^\prime,\tau)$ if the ensemble-averaged potential
$\Phi_0(\bfr)$ is regular, using the same approximations and notation as in
\S\ref{sec:action}.

Assuming that the stars move on their unperturbed orbits in the potential
$\Phi_0$, we may write
\be 
\rho(\bfr,t) = m\sum_{i=1}^N p[\bfr\vert \bfI_i,\bfw_i(t)] =
m\sum_{i=1}^N\sum_{\bfl} p_\bfl(\bfr\vert \bfI_i)
e^{i\bfl\cdot(\bfOmega_it+\bfw_{i0})}.
\ee

Since the initial phases $\bfw_{i0}$ are uniformly distributed, only the terms
with $\bfl={\bf 0}$ survive in the ensemble average; thus the
ensemble-averaged density is 
\be
\rho_0(\bfr)=\langle\rho(\bfr,t)\rangle=(2\pi)^3\int d\bfI F_0(\bfI)p_{\bf
0}(\bfr|\bfI). 
\ee
We may now write
\begin{eqnarray}
\delta\rho(\bfr,t+\tau)\delta\rho(\bfr^\prime,t)  & = &  
m^2\sum_{i,j=1}^N\sum_{\bfl,\bfl^\prime}p_\bfl^*(\bfr\vert \bfI_i)
p_{\bfl^\prime}(\bfr^\prime\vert \bfI_j)e^{-i\bfl\cdot[\bfOmega_i(t+\tau)+
\bfw_{i0}]
+i\bfl^\prime\cdot(\bfOmega_jt+\bfw_{j0})} \nonumber \\
& & -\rho(\bfr,t)\rho_0(\bfr')-
\rho_0(\bfr)\rho(\bfr',t)+\rho_0(\bfr)\rho_0(\bfr').
\end{eqnarray}
Now take an ensemble average; since the phases $\bfw_{i0}$ are randomly
distributed, the average over
$\exp[i(\bfl'\cdot\bfw_{j0}-\bfl\cdot\bfw_{i0})]$ will vanish unless either
(i) $i\not=j$ and $\bfl=\bfl'=0$, or (ii) $i=j$ and $\bfl=\bfl^\prime$.  In
case (i) the sum over $i\not=j$ yields $N(N-1)$ terms, each of which is equal
to $\rho_0(\bfr)\rho_0(\bfr')/N^2$. Thus
\be
\langle\delta\rho(\bfr,t+\tau)\delta\rho(\bfr^\prime,t)\rangle
 =  m^2\left\langle\sum_{i=1}^N\sum_{\bfl}
p_\bfl^*(\bfr\vert \bfI_i)
p_{\bfl}(\bfr^\prime\vert \bfI_i)\exp(-i\bfl\cdot\bfOmega_i\tau)\right\rangle
-{1\over N}\rho_0(\bfr)\rho_0(\bfr').
\ee
We can replace the sum over stars by the
integral over the one-particle \df\ to get 
\begin{eqnarray}
C^{(0)}(\bfr,\bfr^\prime,\tau) & = & \langle
\delta\rho(\bfr,t+\tau)\delta\rho(\bfr^\prime,t)\rangle\nonumber \\
& = & 
(2\pi)^3m\sum_{\bfl}\int d\bfI F_0(\bfI)p_\bfl^\ast(\bfr\vert \bfI)
p_\bfl(\bfr^\prime\vert \bfI)\exp (-i\bfl\cdot\bfOmega\tau)
-{1\over N}\rho_0(\bfr)\rho_0(\bfr');
\label{eq:dampp}
\end{eqnarray}
the superscript ``0'' is a reminder that we have neglected the
self-gravity of the density fluctuations by using the unperturbed
orbits of the stars. This result can also be derived from equations
(\ref{eq:projdef}), (\ref{eq:projop}) and (\ref{eq:cesdens}). 

The dynamic form factor (\ref{eq:fform}) becomes 
\be 
S^{(0)}(\bfr,\bfr^\prime,\omega) =(2\pi)^3 m\sum_\bfl \int d\bfI F_0(\bfI)
p_\bfl^*(\bfr\vert \bfI) p_\bfl (\bfr^\prime \vert \bfI)
\delta(\bfl\cdot\bfOmega-\omega)-{1\over
N}\rho_0(\bfr)\rho_0(\bfr')\delta(\omega). 
\label{eq:fd} 
\ee
Thus the dynamic form factor is determined entirely by resonant stars. 

It is straightforward to calculate the static form
factor (\ref{eq:staform}), starting from equation (\ref{eq:cesdens}):
\begin{eqnarray}
C^{(0)}(\bfr,\bfr^\prime,0) & = & 
\int d\omega S^{(0)}(\bfr,\bfr^\prime,\omega)
 =  m \int d\bfx d\bfv 
F_0(\bfx,\bfv)\delta(\bfx-\bfr)\delta(\bfx-\bfr^\prime)-{1\over
N}\rho_0(\bfr)\rho_0(\bfr') 
\nonumber \\
& = &  m \int d\bfx \rho_0(\bfx)\delta(\bfx-\bfr)\delta(\bfx-\bfr^\prime)
-{1\over N}\rho_0(\bfr)\rho_0(\bfr') \nonumber \\
& = &  m \rho_0(\bfr)\delta(\bfr-\bfr^\prime)
-{1\over N}\rho_0(\bfr)\rho_0(\bfr') .
\label{eq:static}
\end{eqnarray}
This result also follows immediately from equations (\ref{eq:orthog}) and
(\ref{eq:dampp}). 

The correlation function at large times is easily derived from
equation (\ref{eq:dampp}):
\be
C^{(0)}(\bfr,\bfr^\prime,\infty)=
(2\pi)^3m\int d\bfI F_0(\bfI)p_{\bf 0}^\ast(\bfr\vert \bfI)
p_{\bf 0}(\bfr^\prime\vert \bfI)-{1\over N}\rho_0(\bfr)\rho_0(\bfr');
\label{eq:ooooo}
\ee
additional terms would appear if the potential had global resonances such that
$\bfl\cdot\bfOmega=0$ for non-zero $\bfl$ (\cf\ Rauch \& Tremaine 1996).
Thus the modified correlation function defined in equation (\ref{eq:cmod})
takes on the simpler form
\be
\widetilde C^{(0)}(\bfr,\bfr',\tau)=
(2\pi)^3m\sum_{\bfl\not={\bf 0}}\int d\bfI F_0(\bfI)p_\bfl^\ast(\bfr\vert \bfI)
p_\bfl(\bfr^\prime\vert \bfI)\exp (-i\bfl\cdot\bfOmega\tau).
\label{eq:ctill}
\ee

\section{Thermal equilibrium and the fluctuation-dissipation theorem}
\label{sec:thermal}

\noindent
A system with Hamiltonian $mH_0(\bfZ)$ is in thermal equilibrium if the
$N$-particle \df\ has the form $f_0(\bfZ)\propto\exp[-\beta H_0(\bfZ)]$ (in
contrast the term ``equilibrium'' simply denotes that the one-particle \df\ is
a solution of the time-independent collisionless Boltzmann equation).  For the
usual gravitational force, the interaction potential
$m^2\Psi(\bfr)=-Gm^2/|\bfr|$; in this case $f_0(\bfZ)$ diverges exponentially
as $|\bfr_i-\bfr_j|\to0$ and hence a thermal equilibrium state does not
exist. Nevertheless, the useful concept of a thermal equilibrium can be
retained for stellar systems, by modifying the interaction potential in one of
two ways:

\begin{itemize}

\item 
We can eliminate the interaction potential, setting $\Psi(\bfr)=0$, so that the
gravitational potential is determined entirely by the external
potential $\Phi_{\rm ext}(\bfr)$; we then augment $\Phi_{\rm ext}(\bfr)$ by
adding the mean potential field of the unperturbed system.  
This ``mean-field'' approximation
is equivalent to including the self-gravity of the equilibrium stellar system
but neglecting the self-gravity of fluctuations, which is the same as
approximation (4) of \S \ref{sec:intro}. In this case the thermal
equilibrium \df\ is the product of one-particle isothermal \dfs, 
\be
f_0(\bfZ)\propto\Pi_{i=1}^N\exp[-\beta H(\bfz_i)],\quad\hbox{where}\quad
H(\bfz)=\half\bfv^2+\Phi_{\rm ext}(\bfr);
\label{eq:sep}
\ee
in other words the two-particle correlation function vanishes.  This is a
plausible model for galaxies, since (i) the one-particle \df\ 
in galaxies is often approximately isothermal, and (ii) the relaxation times
in galaxies are generally much larger than their age, so significant
correlations between stars have not had time to develop. However, the
mean-field approximation neglects the dynamical effects of the self-gravity of
the perturbed density; 
thus, for example, there is no distinction between the
polarization and response operators and no collective modes.

\item A more realistic approach is to soften the interaction potential to
$m^2\Psi(\bfr)=-Gm^2/(\bfr^2+b^2)^{1/2}$. The softening length $b$ should 
be much
smaller than the size of the system, so the large-scale equilibrium structure
is unaffected; in fact it should also be much less than the
typical interstellar separation so that softening does not affect most
encounters between stars. On the other hand the softening length should be
large enough to suppress the strong correlations between stars that are
present if the gravitational force is unsoftened. To see what is required,
consider an infinite homogeneous stellar system, for which the equilibrium 
two-particle \df\ is
\begin{eqnarray}
p^{(2)}(\bfz_1,\bfz_2) & \propto & 
\exp\left\{-\beta\left[\half v_1^2+\half v_2^2
+m\Psi(\bfr_1-\bfr_2)\right]\right\}
\nonumber \\
& = & p^{(1)}(\bfz_1)p^{(1)}(\bfz_2)
\left[1+g(\bfr_1-\bfr_2)\right],
\end{eqnarray}
where 
\be
p^{(1)}(\bfz)\propto\exp(-\half\beta v^2),\qquad
g(\Delta\bfr)=\exp\left[-\beta m\Psi(\Delta\bfr)\right]
\ee
are the one-particle \df\ and the two-particle correlation
function. We would like the two-particle (and higher order) correlation
functions to be small, so that the $N$-particle \df\ is the product of
one-particle \dfs\ as in equation (\ref{eq:sep}). This requires that  
${\rm max}|\beta m\Psi|$ is small, which in
turn requires $b\gg b_0=Gm/\sigma^2$, where $\sigma^2=1/\beta$ is the
mean-square velocity in one dimension.  Fortunately, it is easy to satisfy 
these conditions in most stellar systems; thus, in
the solar neighborhood we might (for example) choose $b=0.001$ pc, which is
much less than the typical interstellar separation of 1 pc but much greater
than $b_0\sim 10^{-5}$ pc. 

\end{itemize}

\subsection{The fluctuation-dissipation theorem}

\noindent
We examine the relation between the response operator (\ref{eq:uuiioo}) 
and the correlation function (\ref{eq:corrdef}) when the stellar system is in
thermal equilibrium. If $f_0(\bfZ)\propto \exp[-\beta H_0(\bfZ)]$,
\begin{eqnarray}
[f_0,v_{\bfs^\prime}]_{\bfZ_{-\tau}} & = &
-\beta f_0(\bfZ_{-\tau})[H_0,v_{\bfs^\prime}]_{\bfZ_{-\tau}} \nonumber \\
& = & -\beta f_0(\bfZ){d\over d\tau}v_{\bfs^\prime}(\bfZ_{-\tau});
\end{eqnarray}
in the second line, the replacement of the Poisson bracket with a time
derivative follows from Hamilton's equations (\ref{eq:hamdef}), and the
argument of $f_0$ can be changed from $\bfZ_{-\tau}$ to $\bfZ$ because $H_0$
and therefore $f_0$ is invariant along a trajectory.  The response operator
may now be written 
\be 
R_{uv}(\bfs,\bfs^\prime,\tau)={\beta\Theta(\tau)\over m}{d\over d\tau} 
\int d\bfZ u_\bfs(\bfZ)
f_0(\bfZ)v_{\bfs^\prime}(\bfZ_{-\tau})= {\beta\Theta(\tau)\over m}{d\over
d\tau}\int d\bfZ f_0(\bfZ) u_\bfs(\bfZ_\tau) v_{\bfs^\prime}(\bfZ); 
\ee 
in the second equality we have relabeled the variables $\bfZ$ and
$\bfZ_{-\tau}$ as $\bfZ_\tau$ and $\bfZ$ respectively, and then changed the
dummy variable from $\bfZ_\tau$ to $\bfZ$; the Jacobian of the transformation
is unity by Liouville's theorem. Finally, using equation (\ref{eq:corrdef}),
\be 
R_{uv}(\bfs,\bfs^\prime,\tau)={\beta\Theta(\tau)\over m} {\partial\over
\partial\tau}C_{uv}(\bfs,\bfs^\prime,\tau).
\label{eq:wwrrqq}
\ee 
This is the fluctuation-dissipation theorem (Callen \& Welton 1951), which
relates the response operator to the correlation function for any system in
thermal equilibrium. 

If we set $u_\bfs=v_\bfs=\rho(\bfs,t)$, where $\rho(\bfs,t)$ is defined by
equation (\ref{eq:lllkkjj}), then the generalized correlation function $C_{uv}$
and response operator $R_{uv}$ of equations (\ref{eq:corrdef}) and
(\ref{eq:ressdef}) reduce to their analogs $C$ and $R$ defined earlier in
equations (\ref{eq:ccdef}) and (\ref{eq:rdef}). 
Thus the
fluctuation-dissipation theorem implies that 
\be
R(\bfr,\bfr^\prime,\tau)={\beta\Theta(\tau)\over m}
{\partial\over \partial\tau}C(\bfr,\bfr^\prime,\tau),
\label{eq:flucdia} 
\ee 
or
\be
R_o(\bfr,\bfr^\prime,\tau)={\beta\over 2m}
{\partial\over \partial\tau}C(\bfr,\bfr^\prime,\tau)={\beta\over 2m}
{\partial\over \partial\tau}\widetilde C(\bfr,\bfr^\prime,\tau),
\label{eq:flucdiao} 
\ee 
where $R_o$ is defined in equation (\ref{eq:rodef}) and $\widetilde C$ is
defined in equation (\ref{eq:cmod}). The formal analog of this result in
Fourier space is
\be
R_A(\bfr,\bfr^\prime,\omega)=-{i\omega\beta\over 2m}S(\bfr,
\bfr^\prime,\omega); 
\label{eq:flucdiaf} 
\ee the equation is equally valid if $\omega S(\bfr,\bfr',\omega)$ is replaced
by $\omega \widetilde S(\bfr,\bfr',\omega)$ since the difference between the two
functions is proportional to $\omega\delta(\omega)$ which is zero. 

If we suppress the interparticle gravitational interaction (the mean-field
approximation), then the response operator $R(\bfr,\bfr',\tau)$ is replaced by
the polarization operator $P(\bfr,\bfr',\tau)$, and the correlation function
$C(\bfr,\bfr',\tau)$ is replaced by $C^{(0)}(\bfr,\bfr',\tau)$; thus
\be
P(\bfr,\bfr^\prime,\tau)={\beta\Theta(\tau)\over m}{\partial\over \partial\tau}
C^{(0)}(\bfr,\bfr^\prime,\tau). 
\label{eq:flucdiab} 
\ee 
It is straightforward to verify this result when the ensemble-averaged
potential is integrable, using the explicit expressions for
$P(\bfr,\bfr^\prime,\tau)$ and $C^{(0)}(\bfr,\bfr^\prime,\tau)$ in equations
(\ref{eq:dddsss}) and (\ref{eq:dampp}). The analogous equation in Fourier
space is 
\be 
P_A(\bfr,\bfr^\prime,\omega)=-{i\omega\beta\over 2m}S^{(0)}(\bfr,
\bfr^\prime,\omega);
\label{eq:flucdiag}
\ee 
once again, the equation is equally valid if $S^{(0)}$ is replaced by 
$\widetilde S^{(0)}$. 

\section{Correlations in nonisothermal systems 
and the dressed-particle approximation}
\label{sec:dressed}

\noindent
Gilbert (1968) has described collisional relaxation in stellar systems that
are not necessarily in thermal equilibrium, by expanding the exact Liouville
equation in powers of $N^{-1}$. His treatment is based on similar results
derived by Rosenbluth et al. (1957) and Rostoker (1961) for a Coulomb
plasma, and accounts fully for the self-gravity of the medium.

These calculations show that fluctuations in the system can be properly
accounted for---to ${\rm O}(N^{-1})$---by considering an individual test star
and the polarization cloud it induces in the background medium together, as a
``dressed particle''; for a test star with trajectory $\bfr_\ast(t)$ the
density of the bare particle is $\rho_0(\bfr,t)=m\delta[\bfr-\bfr_\star(t)]$,
and the density of the dressed particle is given by
\be
\rho_d(\bfr,\omega)=[\bfI+2\pi\bfR(\omega)\bfPsi]\rho_0(\bfr,\omega) 
= \bfLambda(\omega)\rho_0(\bfr,\omega)
\ee 
(\cf eq. \ref{eq:ldop}).  The test star itself is drawn from the one-particle
\df\ and moves on its unperturbed orbit; thus in this approximation
the only fluctuations in the density of dressed particles are
statistical fluctuations due to the finite number of particles.  The
papers described above effectively show that induced correlations
between dressed particles are ${\rm O}(N^{-2})$ or higher. 

In the dressed-particle approximation, the dynamic form factor $\bfS$ can be
computed from $\bfS^{(0)}$, simply by replacing each particle
by a dressed particle: starting from equation (\ref{eq:rrrnnn}) 
\begin{eqnarray} 
\bfS(\omega)\delta(\omega-\omega^\prime) 
& = & \langle\delta\rho_d(\bfr,\omega)
\delta\rho_d(\bfr^\prime,-\omega^\prime)\rangle
\nonumber \\
& = & \langle \bfLambda(\omega)\delta\rho_0(\omega)
\bfLambda(-\omega')\delta\rho_0(-\omega^\prime)\rangle \nonumber \\
& = & \bfLambda(\omega)\bfS^{(0)}(\omega)
\bfLambda^\dagger(\omega)\delta(\omega-\omega^\prime).
\label{eq:ichi}
\end{eqnarray}
The advantage of this replacement is that $\bfS^{(0)}$ can be computed
directly from the unperturbed orbits of the particles (e.g. eq. \ref{eq:fd},
if the ensemble-averaged potential is regular).

These results are consistent with the two forms of the fluctuation-dissipation
theorem involving the response and polarization operators
(eqs. \ref{eq:flucdiaf} and \ref{eq:flucdiag}), in the following sense. 
The latter equation can be written as $\omega\bfS^{(0)}=(2mi/\beta)\bfP_A$;
substituting this into equation (\ref{eq:ichi}) gives 
\be
\omega\bfS={2mi\over\beta}\bfLambda\bfP_A\bfLambda^\dagger.
\label{eq:ggkkjj}
\ee
Using equation (\ref{eq:rprp}) we can show that 
$\bfLambda \bfP\bfLambda^\dagger=\bfR+2\pi\bfR\bfPsi\bfR^\dagger$, so that
$\bfLambda \bfP_A\bfLambda^\dagger=(\bfLambda
\bfP\bfLambda^\dagger)_A=\bfR_A$; hence equation (\ref{eq:ggkkjj}) simplifies
to
\be
\omega\bfS={2mi\over\beta}\bfR_A,
\ee
which is the same as equation (\ref{eq:flucdiaf}).

Ichimaru (1965) argues that the result (\ref{eq:ichi}) should hold for any
stationary plasma, whether or not in thermal equilibrium, so long as
the relaxation time is much longer than the correlation time; in the context
of stellar systems, the analogous constraint is that the relaxation time
should be much longer than the crossing time, which is equivalent to the
modest requirement that the number of stars $N\gg1$ and entirely
compatible with the results of Gilbert (1968). 

\subsection{Fluctuations and collective modes}

\noindent
One consequence of equation (\ref{eq:ichi}) is that the density fluctuations
described by $\bfS(\omega)$ become very large for frequencies near
singularities of $\bfLambda(z)$ or $\bfR(z)$, i.e. near the frequencies
of collective modes.  To see this more explicitly,
we can expand $\bfR(z)$ in a Laurent series near a collective mode with
complex eigenfrequency $z_0=\omega_0+i\eta$. The resonant part is then
\be
\bfR_{\rm res}(z)={\bfR_{-1}\over z-z_0}.  
\ee 
where $\bfR_{-1}$ is the residual of $\bfR$ at $z_0$.  Consequently, the
dynamic form factor at real frequencies near a weakly damped collective mode
is
\be 
\bfS_{\rm res}(\omega) = {\bfLambda_{-1}\bfS^{(0)}\bfLambda_{-1}^\dagger
\over (\omega-\omega_0)^2+\eta^2} \to
{\pi\bfLambda_{-1}\bfS^{(0)}\bfLambda_{-1}^\dagger\over |\eta|}
\delta(\omega-\omega_0) \qquad\hbox{as $\eta\to0$},
\ee 
where $\bfLambda_{-1}=2\pi\bfR_{-1}\bfPsi$, and we have used equation
(\ref{eq:lbk}).

Note that the level of fluctuations becomes very large as $\eta \tends
0$---that is, as the mode approaches neutral stability. This phenomenon is
analogous to opalescence in the vicinity of a critical
point. Large fluctuations have been observed in numerical simulations of
systems near marginal stability (Ivanov 1992, Weinberg 1993).

\section{Applications of the fluctuation-dissipation theorem}
\label{sec:applications}

\noindent
A useful preliminary calculation is the response of a stellar system in
thermal equilibrium to a static or slowly growing external potential
$\Phi_e(\bfr)$. 
Using equations (\ref{eq:rdef}), (\ref{eq:flucdia})
and (\ref{eq:cmod}) we find
that the density perturbation induced by $\Phi_e$ is
\begin{eqnarray}
\rho_s(\bfr)& = & \int d\bfr^\prime\int_0^\infty d\tau
R(\bfr,\bfr^\prime,\tau)\Phi_e(\bfr^\prime) \nonumber \\ & = &
{\beta\over m}\int d\bfr^\prime\Phi_e(\bfr^\prime)\int_0^\infty d\tau
{\partial\over \partial\tau}\widetilde C(\bfr,\bfr^\prime,\tau) \nonumber \\ & = &
-{\beta\over m}\int d\bfr^\prime\Phi_e(\bfr^\prime)
\widetilde C(\bfr,\bfr^\prime,0).
\label{eq:zerolag}
\end{eqnarray}

If, in addition, we ignore induced fluctuations and approximate
$\widetilde C(\bfr,\bfr^\prime,\tau)$ by 
$\widetilde C^{(0)}(\bfr,\bfr^\prime,\tau)$, and the
ensemble-averaged potential is integrable, we find using equations
(\ref{eq:ctill}) and (\ref{eq:wweezz}) that 
\begin{eqnarray} 
\rho^{(0)}_s(\bfr)& = & -(2\pi)^3\beta\sum_{\bfl\not={\bf 0}}
\int d\bfI F_0(\bfI)p_{\bfl}^\ast(\bfr|\bfI)\Phi_{e,\bfl}(\bfI)\nonumber \\
& = & -\beta \int d\bfv
F_0(\bfr,\bfv)\left[\Phi_e(\bfr)-\langle\Phi_e\rangle\right];
\end{eqnarray}
here $\langle\Phi_e\rangle\equiv\Phi_{e,{\bf 0}}[\bfI(\bfr,\bfv)]$ is the
orbit-averaged potential experienced by the particle passing through the
phase-space point $(\bfr,\bfv)$. The last line is a special case of the
general result (e.g. Lynden-Bell 1969) that the linear response of a stellar
system whose \df\ depends only on energy $E$ to a slowly varying potential is
\be 
f={dF_0(E)\over dE}\left(\Phi-\langle\Phi\rangle\right).  
\ee

\subsection{Dynamical friction on an orbiting body}

\noindent
Consider a body whose center of mass travels on an orbit $\bfr_*(t)$ through a
stellar system, and whose gravitational potential is $\Phi_*(\bfr -\bfr_*)$
(normally $\Phi_*$ is spherically symmetric but this is not necessary for the
derivation; we do, however, neglect changes in the orientation of the body).
We wish to compute the mean force $\overline{\bfF}_1(t)$ exerted on
the body by the response it induces in the stellar system.

The density in the wake, $\rho_s(\bfr,t)$, is given by equation
(\ref{eq:rdef}), and $\overline{\bf F}_1$ is equal and
opposite to the force exerted on the wake by the body, that is, 
\begin{eqnarray}
\overline{\bfF}_1(t)& = & \int
d\bfr\bnabla\Phi_*[\bfr-\bfr_*(t)]\rho_s(\bfr,t)=\int d\bfr d\bfr^\prime
dt^\prime \bnabla\Phi_*[\bfr-\bfr_*(t)]
R(\bfr,\bfr^\prime,t-t^\prime)\Phi_*[\bfr^\prime-\bfr_*(t^\prime)]\nonumber \\
& = & \int d\bfr d\bfr^\prime\bnabla\Phi_*[\bfr-\bfr_*(t)]
\int_0^\infty d\tau R(\bfr,\bfr^\prime,\tau)
\Phi_*[\bfr^\prime-\bfr_*(t-\tau)] .
\label{eq:dragdef}
\end{eqnarray}

For completeness we point out that there is another component of the mean
force on the body, which is not caused by the response it induces in the
system.  The fluctuating force on the body is given by
\be
\delta{\bfF}(t)=\int d\bfr\bnabla\Phi_*[\bfr-\bfr_*(t)]\delta\rho(\bfr,t).
\label{eq:qqqqq}
\ee
Although $\langle\delta\rho(\bfr,t)\rangle=0$, the mean fluctuating force does
not vanish because the orbit $\bfr_*(t)$ is correlated with the fluctuations
$\delta \rho(\bfr,t)$; thus there is an additional contribution to the mean
force $\overline{\bf F}_2=\langle \delta\bfF\rangle$.
 
To compute $\overline{\bf F}_2$, we first compute the perturbation to the
orbit caused by the fluctuating field,
$\bfr_*(t)=\bfr_0(t)+\delta \bfr_*(t)$ where 
\begin{eqnarray}
\delta \bfr_*(t)  =  \int_{-\infty}^t dt^\prime \delta \bfv_*(t^\prime) & = & 
{1\over M} \int_{-\infty}^t dt^\prime \int_{-\infty}^{t^\prime} 
dt^{\prime \prime} \delta \bfF(t^{\prime\prime})
 ={1\over M}\int_0^\infty d\tau \tau\delta \bfF(t-\tau)
\nonumber \\
& = & 
{1\over M}
\int d\bfr^\prime \int_0^\infty d\tau \tau \bnabla^\prime
\Phi_*[\bfr^\prime-\bfr_*(t-\tau)]\delta \rho(\bfr^\prime,t-\tau)
\end{eqnarray}
plus higher-order terms.

Expanding equation (\ref{eq:qqqqq}) and integrating by parts gives, 
\be
\overline{\bfF}_2(t)=\langle \delta{\bfF}(t) \rangle  = 
\int d\bfr \bnabla\Phi_*[\bfr-\bfr_*(t)]
\langle \delta \bfr_*(t)\cdot \bnabla \delta \rho(\bfr,t) \rangle
\ee
where
\begin{eqnarray}
\langle \delta \bfr_*(t)\cdot \bnabla \rho(\bfr,t) \rangle & = &
{1\over M}\int d\bfr^\prime \int_0^\infty  d\tau \tau 
\bnabla^\prime \Phi_*[\bfr^\prime -\bfr_*(t-\tau)]
\cdot \bnabla \langle \delta \rho(\bfr,t)\delta\rho(\bfr^\prime,t-\tau) \rangle
\nonumber \\
& = & 
{1\over M}
\int d\bfr^\prime \int_0^\infty d\tau \tau \bnabla^\prime 
\Phi_*(\bfr^\prime -\bfr_*(t-\tau)]\cdot \bnabla C(\bfr,\bfr^\prime,\tau)
\end{eqnarray}
Thus, 
\be
\overline{\bfF}_2(t)={1\over M} 
\int d\bfr d\bfr^\prime \bnabla\Phi_*[\bfr-\bfr_*(t)]
\int_0^\infty d\tau \tau\bfnabla C(\bfr,\bfr^\prime,\tau)\cdot 
\bfnabla^\prime \Phi_*[\bfr^\prime-\bfr_*(t-\tau)].
\label{eq:ftwo}
\ee 
Note that $\overline{\bfF}_2$ can be formally divergent if
$C(\bfr,\bfr',\infty)$ is non-zero. The reason is that in this case the
time-averaged density is not the same as the ensemble-averaged density, so
that the ``fluctuating'' force can contain a constant component that leads to
a secular change in the orbit. 

The two components of the mean force, $\overline{\bfF}_1$ and
$\overline{\bfF}_2$, have different dependences on the mass of the orbiting
body; in particular, if this mass is much greater than the mass of the stars
in the system, $M\gg m$, then the dominant force is $\overline{\bfF}_1$, which
is greater than $\overline{\bfF}_2$ by O$(M/m)$. In the opposite limit, $M\to
0$ and $\Phi_\star\propto M$, the mean acceleration $\overline{\bf
a}_1=\overline{\bf F}_1/M$ goes to zero as $M$ while the acceleration
$\overline{\bf a}_2$ remains finite.

If the stellar system is in thermal equilibrium, we can use 
the fluctuation-dissipation theorem in the form 
(\ref{eq:flucdia}) to write equation (\ref{eq:dragdef}) as
\be
\overline{\bfF}_1(t)= {\beta\over m}\int d\bfr d\bfr^\prime
\bnabla\Phi_*[\bfr-\bfr_*(t)] \int_0^\infty d\tau{\partial
C(\bfr,\bfr^\prime,\tau)\over\partial\tau}\Phi_*[\bfr^\prime-\bfr_*(t-\tau)].
\label{eq:draga}
\ee 
Now replace $C(\bfr,\bfr',\tau)$ by $\widetilde C(\bfr,\bfr',\tau)=
C(\bfr,\bfr',\tau)-C(\bfr,\bfr',\infty)$ and integrate by parts 
with respect to $\tau$:
\begin{eqnarray}
&\overline{\bfF}_1(t) & = -{\beta\over m}\int d\bfr d\bfr^\prime 
\bnabla\Phi_*[\bfr-\bfr_*(t)]\widetilde C(\bfr,\bfr^\prime,0)
\Phi_*[\bfr^\prime-\bfr_*(t)] \nonumber \\
&{}&-{\beta\over m}\int d\bfr d\bfr^\prime\bnabla\Phi_*[\bfr-\bfr_*(t)]
\int_0^\infty d\tau \widetilde C(\bfr,\bfr^\prime,\tau){\bf v}_*(t-\tau)\cdot
\bnabla\Phi_*[\bfr^\prime-\bfr_*(t-\tau)] \nonumber \\
&\equiv &\overline{\bfF}_s+\overline{\bfF}_d,
\label{eq:dragb}
\end{eqnarray}
where ${\bf v}_*=d\bfr_*/dt$. Using equations (\ref{eq:zerolag}) and
(\ref{eq:qqqqq}) it is easy to show that the first term, $\overline{\bfF}_s$,
is simply the force on the body due to the static response of the stellar
system; in other words this is the force that would result if the body were
fixed in its present position. The second term, $\overline{\bfF}_d$, 
vanishes if $\bfv_*=0$ and corresponds to dynamical friction.
The dynamical friction force can be rewritten as
\be
\overline{F}_{dj}(t)= -{\beta\over m} \int_0^\infty d\tau
v_{\ast i}(t-\tau)\langle\delta F_i(t)\delta F_j(t-\tau)\rangle,
\label{eq:dragc}
\ee 
where the fluctuating forces $\delta F_i(t)$ on the body are computed using
the modified correlation function $\widetilde C(\bfr,\bfr',\tau)$. The result
(\ref{eq:dragc}) holds for any stellar system in thermal equilibrium.

In most calculations of dynamical friction the self-gravity of the wake is
neglected. In this approximation the derivation of equation (\ref{eq:dragc})
remains the same, except that the fluctuation-dissipation theorem is used in
the form (\ref{eq:flucdiab}) and the fluctuating quantities $\delta\rho$ and
$\delta\bfF$ are computed using the correlation function $\widetilde C^{(0)}$,
which neglects interactions between stars.

\subsection{Infinite homogeneous medium}

\noindent
It is instructive to work out the mean force and fluctuating force on a point
mass $M$ traveling through an infinite homogeneous system of stars of mass
$m$. We neglect the self-gravity of the equilibrium system, so that all
objects travel on straight-line orbits at constant velocity; we also neglect
the self-gravity of the response, since otherwise the equilibrium system is
Jeans unstable. Thus there is no distinction between the polarization and
response functions $P$ and $R$, or between the correlation functions $C^{(0)}$
and $C$.

Following equation (\ref{eq:respdens}), the response function may be written
\begin{eqnarray}
R(\bfr,\bfr^\prime,\tau)  & = &  {\Theta(\tau)\over m}
\langle [ \rho(\bfr,t+\tau), \rho(\bfr^\prime,t)] \rangle
\nonumber \\
& = & m\Theta(\tau)\sum_{i j}\langle 
[ \delta(\bfr-\bfr_{i0}-\bfv_{i0}\tau), \delta(\bfr'-\bfr_{j0})] \rangle,
\end{eqnarray}
where the orbit of particle $i$ is $\bfr_i(t+\tau)=\bfr_{i0}+\bfv_{i0}\tau$. 
The Poisson bracket is taken with respect to the phase-space variables 
($\bfr_{i0},\bfv_{i0}$). Thus 
\be
[ \delta(\bfr-\bfr_{i0}-\bfv_{i0}\tau), \delta(\bfr'-\bfr_{j0})]=
-\delta_{ij}\tau\bfnabla \delta[\bfr-\bfr_i(t+\tau)]\cdot \bfnabla^\prime
\delta[\bfr'-\bfr_{j}(t)].
\ee
The response function is then 
\be
R(\bfr,\bfr^\prime,\tau)=-m\Theta(\tau)\tau\sum_i\left\langle 
\bfnabla \delta[\bfr-\bfr_i(t+\tau)]\cdot \bfnabla^\prime
\delta[\bfr-\bfr_{j}(t)]\right\rangle=-{\Theta(\tau)\over m} \tau \bnabla \cdot
\bnabla^\prime C(\bfr,\bfr^\prime,\tau).
\label{eq:relate}
\ee

Integrating equation (\ref{eq:ftwo}) by parts
with respect to $\bfr'$, substituting equation (\ref{eq:relate}), and
comparing to (\ref{eq:dragdef}) shows that in an infinite homogeneous medium
there is a simple relation between the components of the mean force,
\be
\overline{\bfF}_2=\left(m\over M\right)\overline{\bfF}_1.
\ee
Thus
\be
\overline{\bfF}=\overline{\bfF}_1+\overline{\bfF}_2
=\left(1+{m\over M}\right) 
\int d\bfr d\bfr^\prime\int_0^\infty d\tau 
\bnabla\Phi_*[\bfr-\bfr_*(t)]
R(\bfr,\bfr^\prime,\tau)\Phi_*[\bfr^\prime-\bfr_*(t-\tau)].
\ee

To proceed further, we shall assume that the stellar system has a uniform 
number density
$n$ and a Maxwellian velocity distribution with one-dimensional
dispersion $\sigma$---which therefore is isothermal, with
$\beta=1/\sigma^2$. In this case the force correlation function can be shown
to be (\cf Cohen 1975)
\begin{eqnarray}
\lefteqn{\langle F_i(\bfr,t+\tau)F_j(\bfr^\prime,t)\rangle  =  \pi
n(GmM)^2   \times } &&
\nonumber \\
&& \left\{{a_ia_j\over
a^3}\left[\left({3\over u^2}-2\right)\erf(u)-{6\over \pi^{1/2}u}e^{-u^2}\right]
 +  {\delta_{ij}\over
a}\left[\left(2-{1\over u^2}\right)\erf(u)
+{2\over\pi^{1/2}u}e^{-u^2}\right]\right\},
\label{eq:correxp}
\end{eqnarray}
where ${\bf a}=\bfr^\prime-\bfr$, $u=a/(2^{1/2}\sigma|\tau|)$, and
$\erf(x)=2\pi^{-1/2}\int_0^x\exp(-y^2)dy$ is the usual error function. 
The trace of (\ref{eq:correxp}) is simpler,
\be 
\langle {\bf F}(\bfr,t+\tau)\cdot {\bf F}(\bfr^\prime,t)\rangle= {4\pi
n(GmM)^2\over a}\erf(u),
\ee
an expression given by Cohen (1975). 

For a homogeneous system the static induced force $\overline{\bfF}_s$
(eq. \ref{eq:dragb}) vanishes, and we have 
\be
\overline{F}_j=\left(1+{m\over M}\right)\overline{F}_{1j}=
-\left(1+{m\over M}\right)
{v_{\ast i}\over m\sigma^2}\int_{\tau_1}^{\tau_2} d\tau\langle
\delta F_i(\bfr,t+\tau)\delta F_j(\bfr-\bfv_\star\tau,t)\rangle;
\label{eq:ppuuyy}
\ee 
we have changed the limits of integration to $\tau_1>0$ and $\tau_2<\infty$ to
avoid divergences. The integral is easily evaluated to yield
\be
\overline{\bf F}=-{\bfv_\ast}{4\pi G^2Mm(M+m)n\ln\Lambda\over v_\ast^3}
\left[\erf(u)-u\,\erf^\prime(u)\right],
\ee 
where $u=v_\ast/(2^{1/2}\sigma)$ and 
$\Lambda=\tau_2/\tau_1$. This is the standard formula for dynamical
friction in an infinite medium with Maxwellian velocity dispersion
(Chandrasekhar 1943, Binney \& Tremaine 1987), except that the Coulomb
logarithm $\ln\Lambda$ is determined by the limits on the time lag in the
correlation function rather than by limits on the impact parameter. 
Note also that the frictional force induced by the body---which is what we
defined to be dynamical friction in the previous subsection---is actually
$\overline{\bfF}_1$, not $\overline{\bfF}$, which is smaller by a factor
$M/(M+m)$.
 
The effects of relaxation in an infinite homogeneous medium are often
expressed in terms of diffusion coefficients, $\langle \Delta v_j\rangle$ and 
$\langle \Delta v_i\Delta v_j\rangle$, representing the mean and mean-square
changes in velocity per unit time $\Delta t$. These changes are related to the
forces on a point mass $M$ by
\be
M\langle \Delta v_j\rangle=\int_0^{\Delta t}\overline{F}_j(t)dt,\quad
M^2\langle \Delta v_i\Delta v_j\rangle=\int_0^{\Delta t}dt\int_0^{\Delta
t}dt^\prime\langle\delta F_i(t)\delta F_j(t^\prime)\rangle.
\ee
If we assume that the correlation function is negligible for time lag
$>\tau_{\rm max}$, where $\tau_{\rm max}\ll\Delta t$---this is the Markov
approximation of \S 1---then the second of these equations may be written
\be
M^2\langle \Delta v_i\Delta v_j\rangle=
\int_0^{\Delta t}dt\int_{-\tau_{\rm max}}
^{\tau_{\rm max}}\langle\delta F_i(t)\delta F_j(t-\tau)\rangle=
2\int_0^{\Delta t}dt\int_0
^{\tau_{\rm max}}\langle\delta F_i(t)\delta F_j(t-\tau)\rangle,
\ee
where the second equation follows from (\ref{eq:micro}). Equation
(\ref{eq:ppuuyy}) then yields
 \be
\langle \Delta v_j\rangle=-{M+m\over 2 m\sigma^2} v_{\ast i}
\langle \Delta v_i\Delta v_j\rangle,
\label{eq:dvj}
\ee
a result derived already by Chandrasekhar (1943).

\section{Discussion}
\label{sec:discussion}

\noindent
We have described stellar dynamics using the language of statistical physics:
linear response operators, correlation functions, and the
fluctuation-dissipation theorem. It is fair to ask what we have gained from
this rather formal approach, especially since stellar systems do not satisfy
many of the simplifying assumptions commonly used in many-body theory, such as
homogeneity, local forces, and thermodynamic equilibrium. We believe that there
are two main reasons why pursuing this approach is worthwhile. First, these 
techniques have proven to be extremely powerful in other branches of physics,
so it is important to understand to what extent they can be applied to
gravitating $N$-body systems. Second, it is useful to know which features in
stellar systems result from general properties such as causality and
time-reversal symmetry, and which depend on specific approximations used to
make the dynamics tractable. 

An occasional controversy, for example, is the appropriate maximum impact
parameter that should be used in the Coulomb logarithm that enters the
diffusion tensor computed using the local and Markov approximations of \S
\ref{sec:intro}. Chandrasekhar (1942) and Kandrup (1981) advocate terminating
the integration over impact parameters at the typical interstellar separation,
whereas Cohen et al. (1950), followed by most modern authors, argue that the
integration should include all impact parameters up to the characteristic size
of the stellar system (or the Debye length in a plasma). In an $N$-star system
with $N\gg1$, the polarization and response operators depend only on the
one-body \df, which is unchanged if the number of stars is changed so long as
the system mass $Nm$ is conserved. The response operator determines dynamical
friction, and the diffusion tensor is related to dynamical friction through
(\ref{eq:dragc}).  Thus the diffusion tensor is unchanged---except for small
terms whose fractional contribution is O$(m/M)$---if the number of
stars is changed, so long as $Nm$ is conserved. However, this change affects
the interstellar separation; hence the diffusion tensor cannot depend on the
interstellar separation and the effective maximum impact parameter must be of
order the system size.

A central result of this paper is equation (\ref{eq:dragc}), which relates the
dynamical friction force to the force-force correlation function in any
stellar system in thermal equilibrium. This quite general result illuminates
the relation between stochastic and dissipative gravitational forces in
stellar systems. For example, Rauch \& Tremaine (1996) have argued that the
rate of angular momentum relaxation is strongly enhanced in nearly Keplerian
star clusters, such as those found around massive black holes (``resonant
relaxation''). Equation (\ref{eq:dragc}) immediately implies that dynamical
friction is similarly enhanced in such clusters, an effect analyzed by Rauch
and Tremaine and termed ``resonant friction''.

We have not discussed more general issues related to the long-term relaxation of
the one-particle \df, as described by the full collisional
Boltzmann equation or its generalizations. For isothermal systems, the
fluctuation-dissipation theorem already suggests that the Boltzmann collision
integral describing this relaxation must be related to the power
spectrum of potential fluctuations in the background medium.
Indeed, in plasma physics, the corresponding Balescu-Lenard collision term
can be expressed in terms of the fluctuation spectrum determined by the
collisionless plasma (e.g. Lifshitz \& Pitaevskii 1981). Weinberg (1993)
has  derived this collision integral for a model periodic system, 
and argues that fluctuations associated with nearly unstable 
collective modes can strongly enhance the relaxation process.

The concepts in this paper can equally be applied to study artificial
fluctuations and dissipation in numerical methods that approximate stellar
systems, such as self-consistent field codes (Hernquist \& Ostriker 1992),
which approximate the gravitational field by a truncated multipole expansion.

The fluctuation-dissipation theorem in statistical physics also relates
transport coefficients (e.g. electrical and thermal conductivity, diffusion
coefficients) to frequency moments of the fluctuation spectrum. In many cases,
these transport coefficients can be calculated without full knowledge of the
correlation function. Moreover, conjugate transport coefficients are related
by the Onsager relations. Similar relationships may exist for stellar
dynamical systems, although their usefulness remains unclear, especially for
non-isothermal systems.  To study the dynamics of a stellar system described
by a general one-particle \df, it may be possible, for example, to consider
phenomenological response operators that are consistent with the lower-order
moments of the exact response function. These models may offer analytic
insight into the linearized dynamics of the stellar system without computing
the full perturbed distribution function.

\section{Summary}
\label{sec:summary}

\noindent
The main goal of this paper has been to assemble and discuss the general
properties of linear response, dissipation and fluctuations in
stationary stellar systems. Many of the results are not new, having 
already been derived in other arenas of statistical
mechanics---although usually in the context of spatially homogeneous 
systems with short-range forces. Here we summarize our main results.

We have expressed the dynamical response of a stellar system in terms
of a linear response operator $R(\bfr,\bfr^\prime,\tau)$, which
determines the density perturbation induced by an external potential,
$$
\rho_s(\bfr,t)=\int d\bfr^\prime dt^\prime
R(\bfr,\bfr^\prime,t-t^\prime)\Phi_e(\bfr^\prime,t^\prime).
\eqno{(\ref{eq:rdef})}
$$
The response operator is analogous to the conductivity tensor used in
plasma physics; in statistical physics it is sometimes called the
generalized susceptibility (Landau \& Lifshitz 1980).  Many of its
analytic properties in the frequency domain follow from the
requirement that the response must be causal. For example, its
Hermitian and anti-Hermitian parts $\bfR_H(\omega)$ and
$\bfR_A(\omega)$ are related through the Kramers-Kronig relations
(\ref{eq:kramers}).

The anti-Hermitian part of the response operator is associated with
dissipation and dynamical friction; it determines, for example, the 
irreversible work done on the system by an external time-dependent
potential (\S \ref{sec:work}). If the perturbation is periodic, 
$\Phi_e(\bfr,t)=\rm{Re}[\phi_e(\bfr)e^{-i\omega_0 t}]$ the rate 
of irreversible work done is
$$
W=\pi i\omega_0 \int d\bfr d\bfr^\prime
\phi_e^*(\bfr)R_A (\bfr,\bfr^\prime,\omega_0)
\phi_e(\bfr^\prime)=\pi i\omega_0(\phi_e,\bfR_A\phi_e).
\eqno{(\ref{eq:damp})}
$$

For a Hamiltonian system, an exact formal expression for the
response operator is given by the ensemble average of a Poisson
bracket in $6N$-dimensional phase space,
$$
R(\bfr,\bfr^\prime,\tau)
= {\Theta(\tau)\over m}\langle [\rho(\bfr,t+\tau),\rho(\bfr^\prime,t)] \rangle,
\eqno{(\ref{eq:respdens})}
$$
where $\rho(\bfr,t)=m\sum_i \delta[\bfr-\bfr_i(t)]$ is the exact
density distribution of the system.

Equation (\ref{eq:respdens}) generally cannot be explicitly evaluated
for realistic stellar systems (although short-time expansions and sum
rules can be calculated). However, a closely related polarization
operator $P(\bfr,\bfr',\tau)$ is easier to evaluate. The polarization
operator relates the induced density perturbation to the total
(external plus induced) potential,
$$
\rho_s(\bfr,t)=\int d\bfr^\prime dt^\prime
P(\bfr,\bfr^\prime,t-t^\prime)\Phi_t(\bfr^\prime,t^\prime).
\eqno{(\ref{eq:P})}
$$
If the ensemble-averaged potential of the stellar system is
integrable---so that individual stellar orbits have well-defined
actions $\bfI$---the polarization operator is given by
$$
P(\bfr,\bfr^\prime,\omega)= (2\pi)^2 \sum_\bfl \int d\bfI {p_\bfl^*(\bfr\vert
\bfI)p_\bfl(\bfr^\prime \vert \bfI) \over
\bfl\cdot\bfOmega-i\epsilon-\omega}
\,\bfl\cdot {\partial F_0\over\partial \bfI},
\eqno{(\ref{eq:resp})}
$$
where $F_0(\bfI)$ is the one-particle \df, and the functions
$p_\bfl(\bfr\vert \bfI)$ are projection operators onto action space
(eq. \ref{eq:projop}).

The operators $\bfR$ and $\bfP$ are related by a nonlinear equation
(\ref{eq:rprp}). The operator $\bfP$ determines the dispersion
relation for collective modes of the stellar system
(\ref{eq:disp}). The rate of energy loss of a mode to Landau damping
is determined by its anti-Hermitian part, 
$$
W_m=\pi i\omega(\phi_s,\bfP_A\phi_s)=2\eta E_m,
\eqno{(\ref{eq:dampc})}
$$
where $E_m$ and $\eta$ are the energy and damping rate of the mode.
Dissipation occurs through resonant interaction with stellar orbits
commensurate with the mode frequency.

Density fluctuations in a stellar system are characterized by the
correlation function
$$
C(\bfr,\bfr^\prime,\tau)=
\langle\delta\rho(\bfr,t+\tau) 
\delta\rho(\bfr^\prime,t)\rangle.
\eqno{(\ref{eq:ccdef})}
$$
For a stellar system in thermal equilibrium, described 
by an $N$-particle distribution function $f_0(\bfZ) 
\propto \exp[-\beta H_0(\bfZ)]$,
the response function is directly related to the correlation function
by the fluctuation-dissipation theorem, which states that 
$$
R(\bfr,\bfr^\prime,\tau)={\beta\Theta(\tau)\over m}
{\partial\over \partial\tau}C(\bfr,\bfr^\prime,\tau). 
\eqno{(\ref{eq:flucdia})}
$$
The equivalent expression in the frequency domain reads 
$$
R_A(\bfr,\bfr^\prime,\omega)=-{i\omega\beta\over 2m}S(\bfr,
\bfr^\prime,\omega),
\eqno{(\ref{eq:flucdiaf})}
$$
where $S$ is the Fourier transform of the correlation function
(eq. \ref{eq:fform}). Thus dissipational processes in the medium,
determined by $\bfR_A(\omega)$, are directly related to the power
spectrum of density fluctuations.

\begin{acknowledgments}

We thank Martin Weinberg for insights on many of the topics in this paper.  ST
would also like to thank Phil Morrison, who initiated this project by asking
about the relation between dynamical friction and the fluctuation-dissipation
theorem.

This research was supported in part by NSERC, by NASA grant NAG
5-3119 to RWN, and by an Imasco Fellowship to ST. 

\end{acknowledgments}


\begin{thebibliography}{}

\bibitem[Bekenstein \& Maoz (1992)]{bek92} 
Bekenstein, J. D., \& Maoz, E. 1993, ApJ 390, 79

\bibitem[Binney \& Tremaine (1987)]{Bi87} Binney, J. J., \& Tremaine,
S. 1987, Galactic Dynamics (Princeton: Princeton University Press)

\bibitem[Callen \& Welton (1951)]{CW51} Callen, H. B., \& Welton, T. A. 1951,
Phys. Rev. 83, 34

\bibitem[Chandrasekhar (1942)]{Ch42} Chandrasekhar, S. 1942, Principles of
Stellar Dynamics (Chicago: University of Chicago Press).

\bibitem[Chandrasekhar (1943)]{Ch43} Chandrasekhar, S. 1943, ApJ 97, 255

\bibitem[Cohen (1975)]{Co75} Cohen, L. 1975, in Dynamics of Stellar Systems,
ed. A. Hayli (Dordrecht: Reidel), 33 

\bibitem[Cohen et al. (1950)]{CSR50} Cohen, R. S., Spitzer, L., \& Routly,
P. McR. 1950, Phys. Rev. 80, 230

\bibitem[Forster (1975)]{Fo75} Forster, D. 1975, Hydrodynamic
Fluctuations, Broken Symmetry, and Correlation Functions (Reading: Benjamin).

\bibitem[Gilbert (1968)]{Gi68} Gilbert, I. H. 1968, ApJ 152, 1043

\bibitem[Goodman (1988)]{Go88} Goodman, J. 1988, ApJ 329, 612

\bibitem[Hernquist \& Ostriker (1992)]{HO92} Hernquist, L., \& Ostriker,
J. P. 1992, ApJ 386, 375 

\bibitem[Hernquist \& Weinberg (1989)]{HW89} Hernquist, L., \& Weinberg,
M. D. 1989, MNRAS 238, 407

\bibitem[Ichimaru (1965)]{Ich65} Ichimaru, S. 1965, PhysRev B 140, 226

\bibitem[Ivanov (1992)]{Ivan92} Ivanov, A. V. 1992, MNRAS 259, 576

\bibitem[Jeans (1913)]{Je13} Jeans, J. H. 1913, MNRAS 74, 109

\bibitem[Jeans (1916)]{Je16} Jeans, J. H. 1916, MNRAS 76, 552

\bibitem[Kalnajs (1971)]{Ka71} Kalnajs, A. J. 1971, ApJ 166, 275

\bibitem[Kandrup (1981)]{Ka81} Kandrup, H. E. 1981, ApJ 244, 1039

\bibitem[Klimontovich (1986)]{Kl86} Klimontovich, Yu. L. 1986, Statistical
Physics (Harwood: Chur). 

\bibitem[Kubo (1957)]{Ku57} Kubo, R. 1957, J Phys Soc Japan, 12, 570

\bibitem[Kubo et al. (1991)]{Ku91} Kubo, R., Toda, M., \& Hashitsume, N. 1991,
Statistical Physics II, 2nd ed. (Springer: Berlin).

\bibitem[Landau \& Lifshitz (1980)]{La80} Landau, L. D., \& Lifshitz,
E. M. 1980, Statistical Physics, Part I, 3rd ed. (Pergamon: Oxford).

\bibitem[Lifshitz \& Pitaevskii (1981)]{Lp81} Lifshitz,
E. M., \& Pitaevskii 1981, Physical Kinetics. (Pergamon: Oxford).

\bibitem[Lynden-Bell (1969)]{LB69} Lynden-Bell, D. 1969, MNRAS 144, 189

\bibitem[Lynden-Bell \& Kalnajs (1972)]{LBK} Lynden-Bell, D., \& Kalnajs,
A. J. 1972, MNRAS 157, 1

\bibitem[Lynden-Bell \& Wood (1968)]{lbw68} Lynden-Bell, D., \& Wood, R. 1968,
MNRAS 138, 495

\bibitem[Maoz (1993)]{maoz93} Maoz, E. 1993, ApJ 263, 75

\bibitem[Martin (1968)]{ma68} Martin, P. C. 1968, in Probl\`eme \`a N
Corps; Many-Body Physics, eds. C. DeWitt and R. Balian (New York:
Gordon and Breach), 37

\bibitem[Nelson \& Tremaine (1995)]{NT95} Nelson, R. W., \& Tremaine, S. 1995,
MNRAS 275, 897

\bibitem[Palmer (1994)]{Pa94} Palmer, P. L. 1994, Stability of
Collisionless Stellar Systems (Dordrecht: Kluwer)

\bibitem[Rauch \& Tremaine (1996)]{RT96} Rauch, K. P., \& Tremaine, S. 1996,
New Astr 1, 149

\bibitem[Reichl (1980)]{Re80} Reichl, L. E. 1980, A Modern Course in
Statistical Physics (Austin: University of Texas Press).

\bibitem[Rosenbluth, MacDonald \& Judd (1957)]{Ros57} 
Rosenbluth, M. N., MacDonald, W. M., \& Judd, D. L. 1957, Phys Rev 
107, 1

\bibitem[Rostoker (1961)]{Ro61} Rostoker, N. 1961, Nucl Phys 1,
101

\bibitem[Sitenko (1982)]{Si82} Sitenko, A. G. 1982, Fluctuations and
Non-Linear Wave Interactions in Plasmas (Oxford: Pergamon Press).

\bibitem[Spitzer (1987)]{Sp87} Spitzer, L. 1987, Dynamical Evolution of
Globular Clusters (Princeton: Princeton University Press).

\bibitem[Tremaine \& Weinberg (1994)]{TW94} Tremaine, S., \& Weinberg,
M. D. 1984, MNRAS 209, 729

\bibitem[Weinberg (1989)]{Wein89} Weinberg, M. D. 1989, MNRAS 239, 549

\bibitem[Weinberg (1993)]{Wein93} Weinberg, M. D. 1993,  ApJ 410, 543

\end{thebibliography}
\end{document}